\documentclass[preprint,amsmath,amsymb,aps,prb,notitlepage,superscriptaddress]{revtex4-2}

\usepackage{graphicx}
\usepackage[dvipsnames]{xcolor}
\usepackage[colorlinks=False,linkcolor=blue,urlcolor=black,citecolor=blue,bookmarksopen=true]{hyperref}

\newcommand{\figref}[1]{Figure~\ref{#1}}

\begin{document}
\title{Nanoscale spin detection of copper ions using double electron-electron resonance at room temperature}
\author{Kai Zhang}
\affiliation{Department of Physics and Astronomy, University of Pittsburgh, Pittsburgh, PA, 15260, USA}
\affiliation{School of Science and Engineering, The Chinese University of Hong Kong, Shenzhen, Shenzhen 518172, China}
\affiliation{
School of Physical Sciences, University of Science and Technology of China, Hefei 230026, China}
\author{Shreya Ghosh}
\affiliation{Department of Chemistry, University of Pittsburgh, Pittsburgh, PA, 15260, USA}
\author{Sunil Saxena}
\affiliation{Department of Chemistry, University of Pittsburgh, Pittsburgh, PA, 15260, USA}
\author{M. V. Gurudev Dutt}
\email{gurudev.dutt@pitt.edu}
\affiliation{Department of Physics and Astronomy, University of Pittsburgh, Pittsburgh, PA, 15260, USA}
\affiliation{Pittsburgh Quantum Institute, University of Pittsburgh, Pittsburgh, PA, 15260, USA}

\date{\today}

\begin{abstract}
We report the nanoscale spin detection and electron paramagnetic resonance (EPR) spectrum of copper (Cu$^{2+}$) ions via double electron-electron resonance with single spins in diamond at room temperature and low magnetic fields. We measure unexpectedly narrow EPR resonances with linewidths $\sim 2-3$~MHz from copper-chloride molecules dissolved in poly-lysine. We also observe coherent Rabi oscillations and hyperfine splitting from single Cu$^{2+}$ ions, which could be used for dynamic nuclear spin polarization and higher sensitivity of spin detection. We interpret and analyze these observations using both spin hamiltonian modeling of the copper-chloride molecules and numerical simulations of the predicted DEER response, and obtain a sensing volume $\sim (250 \text{nm})^3$. This work will open the door for copper-labeled EPR measurements under ambient conditions in bio-molecules and nano-materials. 
\end{abstract}
\maketitle

\section{Introduction}
Electron paramagnetic resonance (EPR) spectroscopy has emerged as a versatile technique with uses in different fields of science and engineering.  Spectral characteristics of EPR signals such as intensity, lineshape and position allows one to extract information on the properties of a magnetic system and its local environment. Such information has had much recent use in the measurement of protein structure and dynamics~\cite{Hubbell:2013aa,Ubbink:2002aa}, identification of point defects in semiconductors~\cite{Feher:1959aa,Watkins:1998aa,Loubser:1978aa}, paramagnetic reaction intermediates~\cite{Davydon:2009aa}, and characterization of photochemical reactions~\cite{Niklas:2017aa,Shaakov:2003aa,Britt:2000aa}, to name a few.

Of the several widespread sensors of the EPR signal, the nitrogen-vacancy (NV) center is of significant interest in quantum sensing~\cite{Taylor08,Maze08,Balasub08,Degen2017,Barry2020,Casola2018}. Diamond quantum magnetometers with NV centers have been shown to have excellent sensitivity and resolution~\cite{Taylor08,Maze08,Balasub08,Staudacher2013,Mamin2013}. These properties are enabled by the fact that the NV center is an atomic-scale defect with long spin relaxation and spin coherence times even at room temperature due to the isolation from the environment~\cite{Wrachtrup06rev,Jelezko04a,Jelezko04b,Childress06b,Balasub09}. The NV center can also be probed optically which makes it ideal for non-invasive detection and imaging of many biological, chemical and nano-material applications~\cite{Doherty2013,Degen2017,Barry2020,Casola2018,FShi:2018aa,Fortman:2019aa}, including recent magnetic resonance detection of single proteins~\cite{Shi:2015aa,Lovchinsky:2016bb}. 

EPR based measurements in biophysics most commonly involve labeling with nitroxides~\cite{Hubbell:2013aa}. Likewise, for single spin detection using NV centers in diamonds, nitroxide spin labels have been widely used~\cite{Shi:2015aa,Akiel:2016aa,Akiel:2017aa}. Recently, the developments in spin-labeling techniques have led to alternative labels that use other organic radicals or paramagnetic metal ions~\cite{Ubbink:2002aa}. Particularly, spin labeling using Cu$^{2+}$ has gained importance, because Cu$^{2+}$ is one of the most abundant cofactor metals in proteins, binding to several metalloproteins.  Recently, Cu$^{2+}$ bound to two strategically placed histidine residues, also known as the Cu$^{2+}$-dHis motif, have provided several significant advancements towards characterization of protein conformations~\cite{Cunningham:2015aa,Sameach:2019aa,Lawless:2018aa,Singewald:2020aa}. The Cu$^{2+}$-dHis motif provides nearly five-fold narrower distances than nitroxides, potentially enhancing the resolution of EPR distance methods~\cite{Cunningham:2015aa,Ghosh:2018aa}. Furthermore, the rigidity of the label enables orientation selectivity that has been shown to correlate with the protein subunit conformation~\cite{Jarvi:2018aa,Bogetti:2020aa}. An important future direction for this research is the development of methodology that can detect the Cu$^{2+}$ EPR signal at room temperatures with high sensitivity. Cu$^{2+}$ ions also have larger anisotropy in g-factors, and larger hyperfine coupling strengths, which leads to larger splitting in the spectrum. This could be an advantage for selective driving of the hyperfine transition lines, commonly used for dynamic nuclear polarization and advanced quantum sensing techniques~\cite{Lovchinsky:2016bb}. In this work, we explore the potential of NV-centers to detect the EPR signal from Cu$^{2+}$. 

\section{Materials and Methods}
\label{sec:methods}
\begin{figure}[htb]
\includegraphics[scale=1.0]{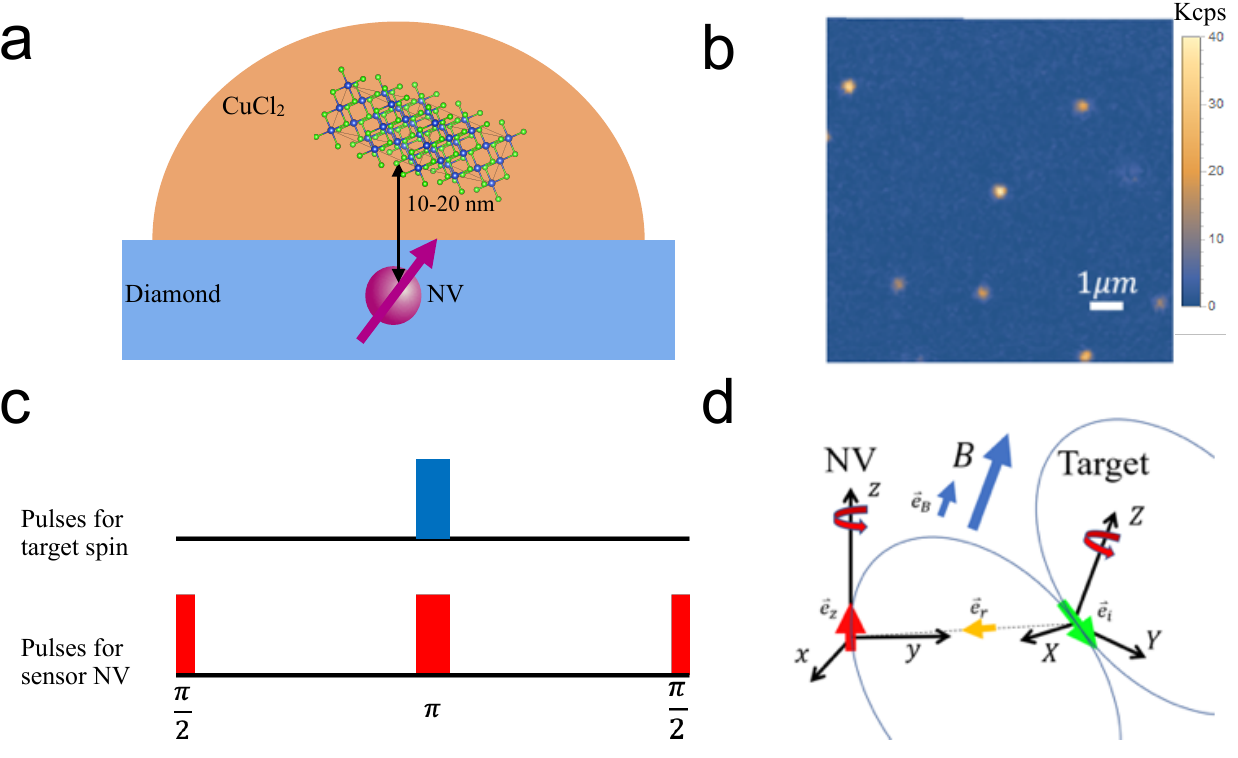}
\caption{(a) Illustration of quantum sensing of CuCl$_2$ molecules by an NV center which is $\sim 10 - 20$ nm deep under the diamond surface. The orange hemispherical region (not to scale) represents the sensing volume of the NV center. (b) Confocal image of implanted NV centers. The diameter of the gaussian bright spots, representing single NV centers, is 300 nm. (c) Typical experimental sequence used in double electron-electron (DEER) resonance with NV centers. The pulse sequence for the NV center is a spin echo sequence which is used for detecting AC magnetic fields. When the pulse for the target electron spin is applied, the resultant change in the target spin causes an AC magnetic field that is detected by the NV center. (d) Illustration of the theoretical model used for the derivation of the DEER signal from single target spins on the NV center. Separate coordinate systems are used for the NV and target spins in the lab frame, and the unit vectors $\hat{e}_B, \hat{e}_i, \hat{e}_r$ are explained in the main text. }
\label{fig:1}
\end{figure}

\figref{fig:1}(a) is a schematic illustration of the setup, with NV centers located $\sim 10 - 20$ nm away from the Cu-labeled molecules, while microwaves and optical illumination are used to manipulate and detect the spin of the NV center. The sensor NV is usually driven by a resonant spin-echo pulse sequence to zero out effects of slowly fluctuating magnetic fields including the spin bath of $^{13}$C nuclear spins, as shown in \figref{fig:1}(c). The spin-echo pulse effectively acts as a filter for only those fields that fluctuate at frequencies $\sim 1/\tau$ where $2\tau$ is the length of the pulse sequence~\cite{Taylor08,Maze08,Balasub08}. By placing another pulse to drive spin transitions in the target molecule right at the mid-point we cause precession of the target spin, and thereby a fluctuating magnetic field at the frequency of the spin-echo sequence which results in a dip in the NV spin echo signal~\cite{Grotz2011,Mamin2012:bb,Sushkov2014a,Grinolds2014,Shi:2015aa}. This DEER pulse can be scanned in frequency and time to obtain information about the target spin. As we will see later, for experimental reasons, we typically use two DEER pulses: one placed after the $\pi/2$ and the other after the $\pi$ pulse of the sensor NV pulse sequence, but the essential idea remains the same. 

Our diamond samples are $\langle 100 \rangle$-cut CVD electronic grade diamond (Element Six) with specified low concentration of native $^{14}$N ($<1$ ppb). The samples were implanted with $^{15}$N ions at 14 keV energy and dose of $1 \times 10^9$/cm$^2$ at a 7$^\circ$ angle of incidence (INNOViON). Our SRIM and TRIM simulations show average implantation depth to be $h \sim 20$~nm. The samples were then annealed in a tube furnace with a forming gas atmosphere (N$_2$ and H$_2$, 10 mTorr pressure) at 1000$^\circ$C for two hours. The graphitization of diamond surface during the annealing is removed by reflux in a tri-acid mixture of 1:1:1 sulfuric, nitric, and perchloric acid for an hour. A small amount of water solution containing 100 nmol of CuCl$_2$ is mixed into 1 mL water solution with 0.01\%  concentration of poly-L-lysine. A small droplet ($\sim 5 \, \mu$L) of the mixed solution with Cu$^{2+}$ target spins is dropped on the diamond surface with the implanted NV centers. After drying out, the target spins with poly-L-lysine residuals holding them in position are left on the diamond surface. 

The samples are placed in our confocal microscopy setup for quantum magnetometry that is implemented with scanning sample mechanism and is described elsewhere~\cite{KaiThesis2018}. The diamond is placed on top of a coverslip with fabricated coplanar waveguide, which is fabricated with photo-lithography and metal deposition. A tiny amount of immersion oil is placed on the cover slip prior to placing the diamond to fill the air gap between the diamond and coverslip and provide higher resolution. A coverslip-corrected oil immersion objective (Olympus UPLFLN100XO2) is used to focus on the diamond surface and locate the NV centers. The coverslip is glued to a sample holder mounted to a 3-axis piezo nanopositioning stage (MadCityLabs Nano-LP100). When we scan the position of the sample mount, a fluorescence image of implanted surface NV centers is obtained as shown in \figref{fig:1}(b). Typical lateral resolution of our confocal microscope is $\sim 0.3 \, \mu$m, while longitudinal resolution is $\sim 1 \, \mu$m. We have observed typical saturated photon count rates $\sim 1.3 \times 10^{5}$ counts per second from a single NV center, because of the high NA (numerical aperture) of the oil-immersion objective.

In DEER experiments, the microwave pulses to drive NV centers and target spins are at different frequencies, because the zero-field splitting of NV center makes the Larmor frequency of NV center different from typical electron spins with $g \sim 2$. The microwave circuit is implemented using two separate microwave synthesizers (PTS 3200, Windfreak Tech SynthUSBII) with independent control and separate microwave switches triggered by independent channels of an arbitrary waveform generator (Tek AWG520). The two separate microwaves are combined and amplified before being delivered to the NV center. The coplanar waveguide fabricated on the coverslip is connected to our microwave circuit with soldered microwave SMA connector at one end, leaving the other end of the waveguide open. Our setup can achieve fairly high and stable Rabi frequency ($>25$ MHz) for NV centers due to high efficiency of coplanar waveguide, corresponding to a $\pi$-pulse length of $\sim 20$ ns. To generate a constant magnetic bias field and align it with the N-V axis of the chosen NV center, a permanent magnet is held by two rotation mounts hanging from a fixed beam. The two rotation mounts are responsible for the azimuthal angle and polar angle independently, resulting in controllable orientation of the magnetic field in all possible directions in space. 

\section{Theory}
\subsection{Single Electron Spin DEER}
\label{sec:singleDEER}
Our simplified physical model for DEER assumes that the target spin is a single electron spin near the NV center and the only interaction between the two arises from a magnetic dipole-dipole interaction. \figref{fig:1}(d) shows the coordinate axes for the two rotating frames of the system of the sensor NV and the target spin in the lab frame. The Hamiltonian of this interaction can be written as,
\begin{equation}
H = - \frac{\mu_0}{4 \pi r^3}[3 (\vec{\mu}_1 \cdot \hat{e}_r)(\vec{\mu}_2 \cdot \hat{e}_r) - \vec{\mu}_1 \cdot \vec{\mu}_2]
\end{equation}
where $r$ is the distance between the two magnetic dipoles $\vec{\mu}_1$ and $\vec{\mu}_2$, and $\hat{e}_r$ is the unit vector indicating the direction of the displacement between the magnetic dipoles $\vec{\mu}_1$ and $\vec{\mu}_2$. In our DEER experiments, $\vec{\mu}_1$ is the target spin and $\vec{\mu}_2$ is the NV center. Thus the Hamiltonian can also be written as,
\begin{equation}
H = -  \vec{B}_{12} \cdot \vec{\mu}_{2} 
\end{equation}
where 
\begin{equation}
\vec{B}_{12} = \frac{\mu_0}{4 \pi r^3}[3 (\vec{\mu}_{1} \cdot \hat{e}_r)\hat{e}_r - \vec{\mu}_{1} ]
\end{equation}
is the magnetic field at the NV location due to the copper spin $\vec{\mu}_{1}$. For small magnetic fields, the NV center however is only sensitive to the $z$-component of this field, and therefore we can write the effective field as,
\begin{equation}
B_z = \hat{e}_z \cdot \vec{B}_{12}  =  \frac{\mu_0}{4 \pi r^3}[3 (\vec{\mu}_{1} \cdot \hat{e}_r)(\hat{e}_r \cdot \hat{e}_z)- \vec{\mu}_{1} \cdot \hat{e}_z ]
\end{equation}
After introducing the target spin magnetic dipole $\vec{\mu}_1 = \tfrac{\gamma_e \hbar}{2} \hat{e}_1$, where $\gamma_e$ is the gyromagnetic ratio for electron spin, this can be transformed to the simplified form,
\begin{equation}
B_z = \hat{e}_1 \cdot \bigl\{ \frac{\mu_0 \gamma_e \hbar}{8 \pi r^3}[3 (\hat{e}_r \cdot \hat{e}_z) \hat{e}_r - \hat{e}_z] \bigr\} = \lambda_{i} \hat{e}_1 \cdot \hat{e}_{i}
\end{equation}
where $\lambda_{i}$ and $\hat{e}_{i}$ are the magnitude and direction of the vector representing the dipolar magnetic field on the sensor NV spin. If we use spherical polar coordinates for the unit displacement vector $\hat{e}_r = (\sin \theta_{r} \cos \phi_{r},\sin \theta_{r} \sin \phi_{r},\cos \theta_{r})$ we get,
\begin{align}
    \lambda_i &= \frac{\mu_0 \gamma_e \hbar \sqrt{3 \cos^2 \theta_{r} +1}}{8 \pi r^3} \\
    \hat{e}_{i} &= \frac{[3(\hat{e}_r \cdot \hat{e}_z) \hat{e}_r - \hat{e}_z]}{\sqrt{3 \cos^2 \theta_{r} +1}}
\end{align}

Now we can calculate the effect of this field on the sensor NV during the spin echo sequence, neglecting all other magnetic fields from spin bath except from the target spin. We also apply a drive DEER pulse in the middle of the spin-echo as shown in \figref{fig:1}(c), with Rabi frequency $\Omega$, detuning $\Delta$ from the resonance of the target spin, and with length $t_p$. In that case, we can show the net phase accumulated during the sequence is,
\begin{equation}
\phi  = \gamma_e \tau \lambda (\hat{e}_B \cdot \hat{e}_{i}) [ (R_B (\phi_r) R_{\vec{a}} (\alpha) \hat{e}_1 - \hat{e}_1). \hat{e}_B]
\end{equation}

\begin{figure}[htb]
\includegraphics[width=0.7\textwidth]{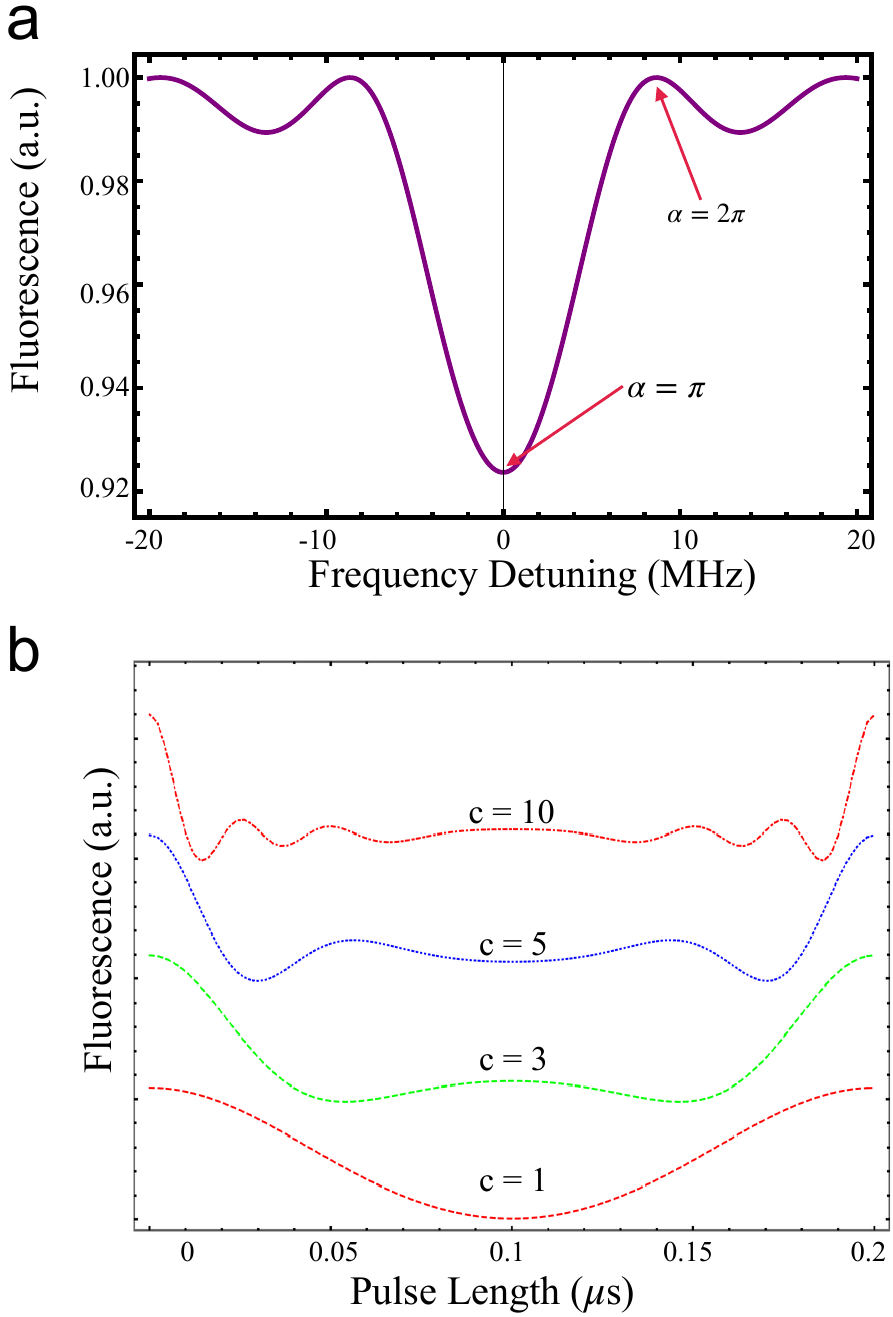}
\caption{(a) Calculation of the DEER spectrum from a single target spin using numerical integration of $f(\Delta)$ from \eqref{eq:ftheory}. The pulse length of the target spin driving field is designed to make a $\pi$ pulse when resonant ($\Omega = 5$ MHz, $t_p = 0.1 \, \mu$s). (b) Numerical simulation of the expected DEER Rabi experiment signals from a single target spin for different values of the prefactor $c = 10,5,3,1$ with fixed $\Omega = 5$ MHz and $\Delta = 0$. The presence of oscillations within one collapse-revival time of the Rabi signal is seen when $c \gg 1$. Plots for different prefactors $c$ are offset for clarity.}
\label{fig:2}
\end{figure}

Here $\hat{e}_B$ is a unit vector along the direction of the external magnetic field applied to the sample, $\tau$ is the delay betwen pulses, $\vec{a} \parallel (\Omega, 0, \Delta) $ represents the drive field vector in the rotating frame, $\alpha = t_p \sqrt{\Omega^2 + \Delta^2} $ is the angle of rotation, $R_B$ and $R_a$ represent rotation operators along the $\hat{e}_B$ and $\vec{a}$ directions. $\phi_r$ represents the random phase between the drive DEER pulse and the initial microwave $\pi/2$-pulse of the NV sequence, and therefore the applied pulse causes a random nutation dependent on the target spins position and resonant frequency. The random orientation of the $\hat{e}_1$ vector arises from the fact that the target spin is not initialized to a fixed state. The measured fluoresence signal of the NV center is therefore an ensemble average over all possible directions of $\hat{e}_1$ and angles $\phi_r$:
\begin{widetext}
\begin{equation}
f(\Omega,\Delta,t_p) = \int_0^{2\pi} d\phi_r \int_{-1}^{1} d(\cos \theta_1) \int_0^{2 \pi} d \phi_1 \cos \bigl\{ c [ (R_B(\phi_r) R_{\vec{a}}(\alpha) \hat{e}_1- \hat{e}_1 ) .\hat{e}_B] \bigr\} 
\label{eq:ftheory}
\end{equation}
\end{widetext}
where $\theta_1$ and $\phi_1$ are the polar angle and azimuthal angle of $\hat{e}_1$, and $c = \gamma_e \tau \lambda (\hat{e}_B \cdot \hat{e}_1) $ is the dimensionless prefactor. The fluorescence is a function of the  Rabi frequency $\Omega$, detuning $\Delta$ , and the pulse length $t_p$ of the DEER pulse driving the target spin. The dependence is folded into the rotation operator $R_{\vec{a}}(\alpha)$. Unfortunately, there is no analytical solution for this integral, but we can numerically integrate \eqref{eq:ftheory} to calculate $f(\Delta)$ or $f(t_p)$ while fixing the other parameters. \figref{fig:2}(a) shows the prediction for the cases where $\Delta$, the detuning frequency of the driving microwave pulse, is varied while keeping $\Omega$ and $t_p$ fixed; while \figref{fig:2}(b) calculates the theoretical prediction for  the situation where the $t_p$, the length of the DEER pulse is varied.

In \figref{fig:2}(a), the parameters $\Omega$ and $t_p$ are chosen so that the driving DEER pulse causes a $\pi$ rotation of the target spin when the frequency of the driving field is on resonance ($\Delta = 0$). The first revival of the flurorescence signal happens when the frequency detuning satisfies $\alpha = t_p \sqrt{\Omega^2 + \Delta^2}  = 2 \pi$. The detuning at which this revival happens is therefore given by,
\begin{equation}
\Delta_R = \sqrt{\frac{4 \pi^2}{t_p^2} - \Omega^2}
\end{equation}
\figref{fig:2}(b) also shows the calculation for the DEER signal when the pulse length $t_p$ is varied, which we call the DEER Rabi experiment. The frequency of the target spin drive pulse is assumed to be on resonance $\Delta = 0$. Since the rotation operator $R_{\vec{a}}(\alpha)$ is periodic in the time $t_p$, so is the function $f(t_p)$, and so our simulation only covers one time period of this rotation. The effect of the pre-factor $c$ is clearly seen in the figure, leading to dramatic changes in the shape of the function $f(t_p)$. For the situation when $c > 3$, there is an obvious deviation from sinusoidal behavior. In the strong coupling regime (e.g.$c = 10$), the fluorescence signal has clear oscillations within one revival-period. It is worth noting that these oscillations were mistakenly thought to be Rabi oscillations of the target spin in Ref.~\cite{Mamin2012:bb}, and thus led to incorrect $2 \pi$ pulse length for the target spin. We can also see that the pre-factor needs to be $c \sim 1$ at least to achieve reasonable contrast of the fluorescence signal. In our experiments, the typical time of the spin echo sequence $ \tau = 6 \, \mu$s, thus the $c = 1$ condition can be met when the depth of the NV center is $h \sim 10$~nm; while $h \sim 4$~nm for the $c = 10$ condition to be met.

\subsection{Ensemble DEER}
\label{sec:ensembleDEER}

The theory of DEER experiment with single electron spins can also be extended to multiple electron spins. Assuming the interaction between the target spins is negligible compared to their interaction with the NV center, the magnetic field at the NV center location is given by,
\begin{equation}
B_z = \sum_k \lambda_k \hat{e}_k \cdot \hat{e}_{i,k}
\end{equation}
where $k = 1, 2,\ldots,n$, and $n$ is the total number of electrons. The net phase imparted during the spin-echo sequence is then,
\begin{equation}
\phi = \gamma_e \tau \sum_k \lambda_k (\hat{e}_B \cdot \hat{e}_{i,k}) [ (R_B(\phi_r) R_{\hat{a}} (\alpha) \hat{e}_k - \hat{e}_k) \cdot e_B]
\end{equation}
And the measured fluorescence signal $f \propto \langle \cos \phi \rangle_{1,2,\ldots,k,\phi_r}$ where $\langle \ldots \rangle_{1, \ldots,k,\phi_r}$ is the average over all the direction of the target spins and the random phase $\phi_r$. The result is also based on the assumption that the gradient of the bias magnetic field and the microwave driving fields is negligible on the length-scales separating the different target spins.

\begin{figure}[htb]
\includegraphics[width=\textwidth]{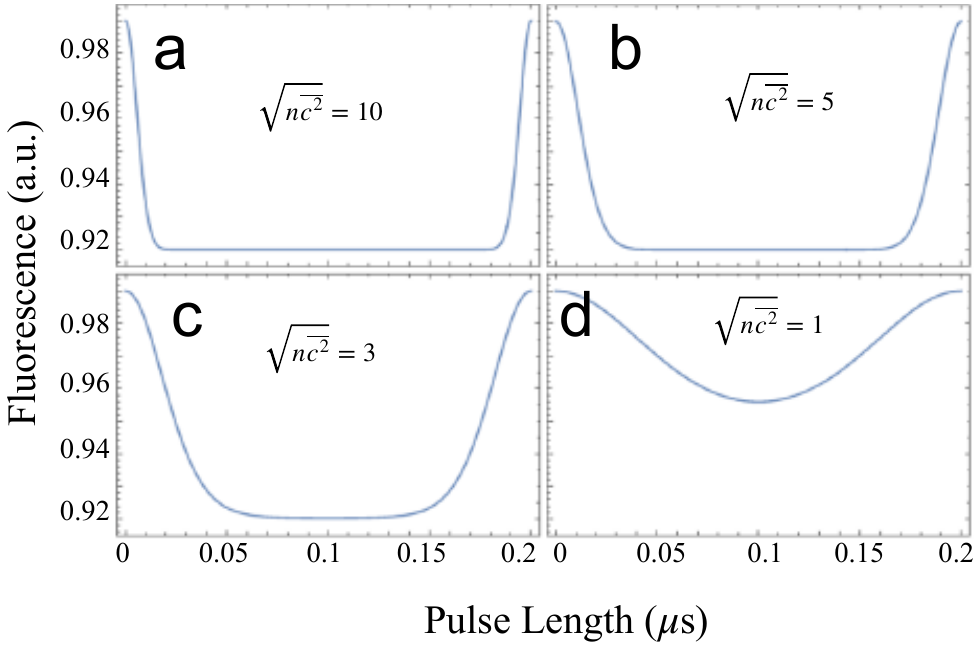}
\caption{(a) - (d) Numerical simulation of the DEER Rabi experiment from an ensemble of target spins for values of the prefactor (denoting strength of coupling) $\sqrt{n\overline{c^2}} = 10 \, , \, 5, \, 3,\, 1$ respectively, with fixed values of $\Omega = 5 $ MHz and $\Delta = 0$. Even when the prefactor becomes much larger than 1, there are no oscillations. }
\label{fig:3}
\end{figure}

For a small number of target spins nearby with different $\lambda_k$ and $\hat{e}_{i,k}$, the probability has to be carefully evaluated for each spin and averaged. However, when there is a truly large number of nearby target spins with small prefactors representing the strength of the interaction with each spin, the net phase is the sum over many independent random contributions. As such, by the central limit theorem, the net phase $\phi$ will tend towards a normal distribtution with mean value $0$ and variance $\sigma_\phi^2$. Thus we can now make sure of the well known result that $ \langle \cos \phi \rangle = \exp (- \sigma_\phi^2/2)$, when $\phi$ is normal distributed. It can also be shown from the expression for the $\phi$ that the variance will simply be the sum of the individual variances which is,
\begin{equation}
\sigma_\phi^2 = n \bar{c}^2 \frac{4 \Omega^2}{3(\Omega^2 + \Delta^2)} \sin^2 (\pi t_p \sqrt{\Omega^2 + \Delta^2})
\end{equation}
where the prefactors $c_k^2$ have been replaced by their average value $\bar{c}^2$. Thus the final result for the ensemble DEER signal is,
\begin{equation}
f_{}(\Omega,\Delta,t_p) = \exp \biggl[ -\frac{2 n \bar{c}^2  \Omega^2}{3(\Omega^2 + \Delta^2)} \sin^2 (\pi t_p \sqrt{\Omega^2 + \Delta^2}) \biggr]
\end{equation}
\figref{fig:3} shows the simulation of the DEER Rabi experiment on an ensemble of target spins. In contrast to \figref{fig:2}(b), the fast oscillations of the signal with pulse length $t_p$ disappear even when there is strong coupling to the ensemble of target spins. As a result, the oscillations within one period of the DEER Rabi experiment could potentially be used to distinguish single versus ensemble of target spins for strong coupling strengths. However the simulation result for $c=1$ in \figref{fig:2}(b) and for $n \bar{c}^2 =1$ in \figref{fig:3}(d) are sufficiently similar that it may not be possible to distinguish the two cases for weak coupling strengths, based on DEER Rabi data alone.

\section{Results} 
\subsection{Frequency spectrum of DEER}
\label{sec:expdeerspectrum}
\begin{figure}[htb]
\includegraphics[width=0.8\textwidth]{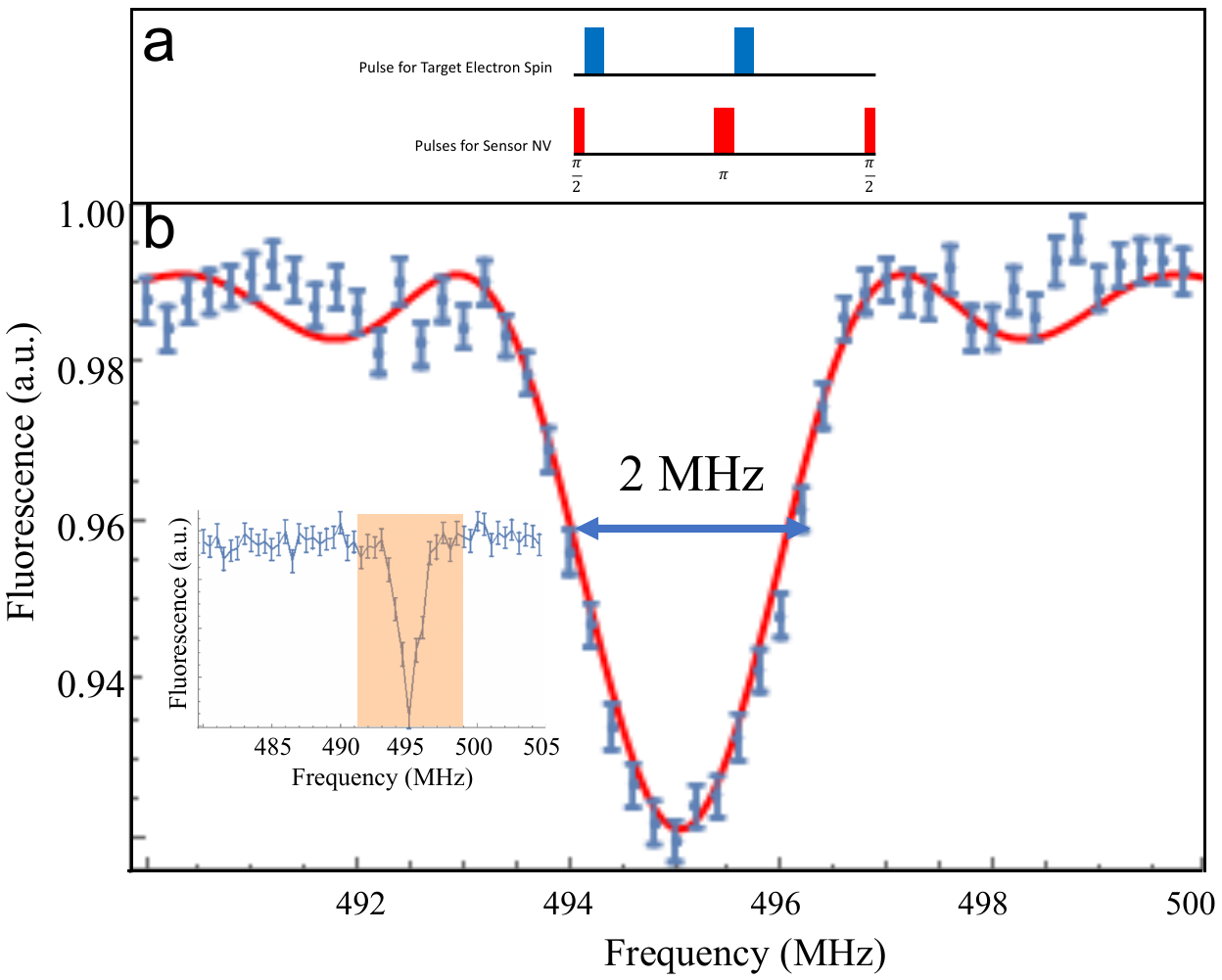}
\caption{(a) Modified experimental sequence for DEER. The target spin driving pulse in the middle is slightly shifted after the sensor $\pi$ pulse, and an additional target spin driving pulse is added after the sensor $\pi/2$ pulse to balance any artifacts from simultaneous propagation of microwaves in the waveguide structure~\cite{Mamin2012:bb}. (b) Experimentally obtained DEER spectrum of the target spin near NV1 for magnetic field $B = (114^{\pm2.5}, 0, 163^{\pm2.5})$ Gauss (inferred from NV ODMR data) where the z-axis (NV axis) is along the [111] direction. The pulsewidth for the target spin used was 100 ns, the $\pi$-pulse width for the NV center is 20 ns, and the spin-echo time $\tau = 6.6 \, \mu$s which is the revival time of the $^{13}$C nuclear spin bath. Each data point represents signal accumulated from $2.5 \times 10^{7}$ repetitions of the pulse sequence. Error bars are obtained from the Poisson statistics of the photon counts and represent $\pm 1 \sigma$ confidence intervals. The solid line is a fit to the data with a sinc-squared function which can be derived for perfectly resonant $\pi$ pulse driving the target spin, since probability of transition is then given by $ P = \frac{\pi^2}{4} \,\text{sinc}^2 (\frac{\pi \sqrt{\Omega^2 + \Delta^2}}{2 \Omega})$. (inset) Frequency scan over a wider range of the same DEER resonance from NV1. The shaded region shown has been zoomed in for the main plot. }
\label{fig:4}
\end{figure}

As introduced in Section~\ref{sec:methods} and shown in \figref{fig:1}(c), the DEER drive pulse is normally applied at the same time as the $\pi$ pulse applied to the sensor NV during the spin-echo sequence. However, because we observed large artifacts in the signal caused by having two different microwave pulses with different frequencies at the same time, the sequence was slightly modified~\cite{Mamin2012:bb}. A pair of balanced drive pulses is applied right at the beginning and the middle of the sensor NV echo sequence, right after the $\pi/2$ and $\pi$ pulses respectively, as shown in \figref{fig:4}(a).  The sequence length $\tau$ is chosen to be at the revival caused by the $^{13}$C Larmor frequency~\cite{Childress06b,Maze08}. The frequency of the DEER pulse is scanned while the pulse length is kept fixed at $100$~ns. After finding the approximate resonance (see inset to \figref{fig:4}), we optimize the pulse length and then carry out a fine scan as shown in the main panel of \figref{fig:4}(b). We note the linewidth (FWHM) of this resonance $\Delta \nu = 2$ MHz, which is much narrower than expected. This data was taken with an implanted NV center (NV1) that had coherence times $T_2 \sim 1 \, \mu$s, measured using the spin-echo sequence. 

\begin{figure}[htb]
\includegraphics[width=\textwidth]{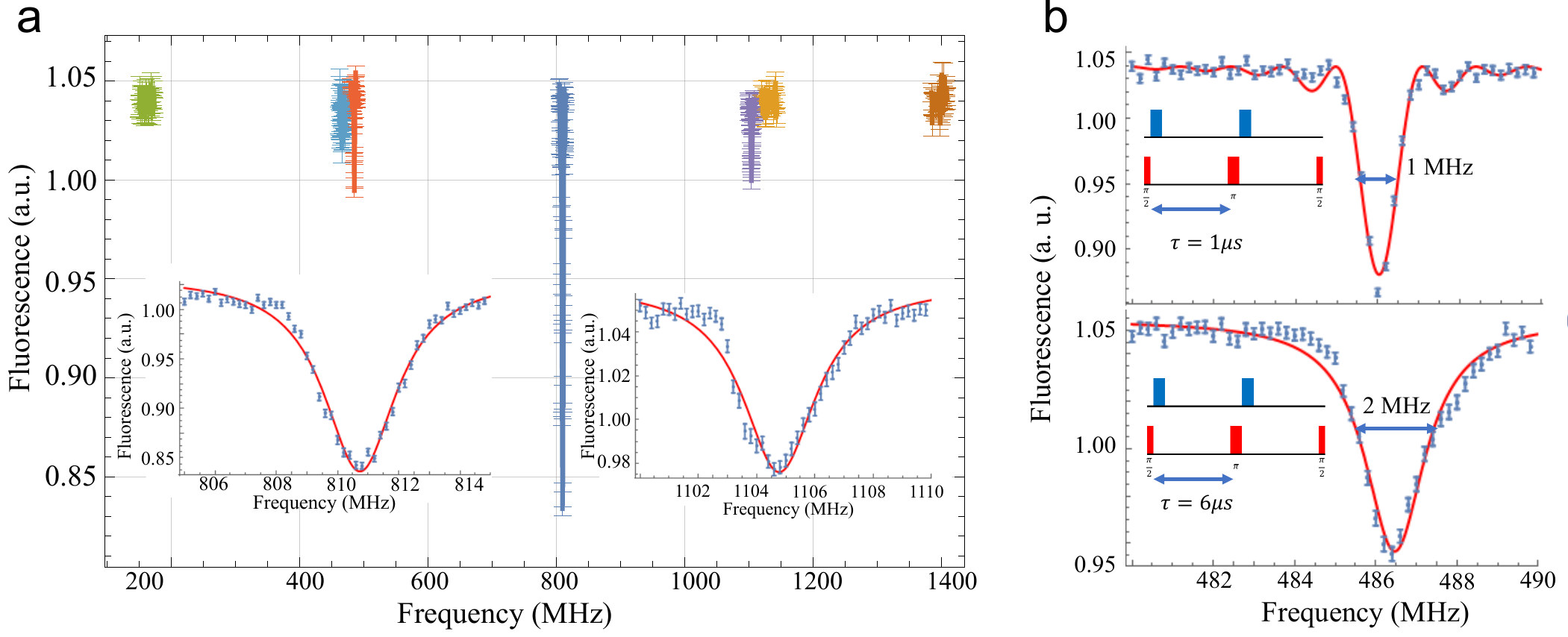}
\caption{(a) Experimental DEER spectrum from NV2 using $\tau = 6 \, \mu$s spin-echo sequences while scanning the drive pulse frequency over other frequency ranges. The magnetic field used was $B = (114^{\pm2.5}, 0, 163^{\pm2.5})$ Gauss and remains nearly constant in all the remaining DEER experiments.  The insets show zoomed in scans near the resonances at $f_{deer} = 810$ MHz and $f_{deer} = 1104.5$ Mhz.  The drive pulse width for the left inset is 130 ns, while the drive pulse width for the right inset was 380 ns. Red curves are Lorentzian fits. (b)  Experimental DEER spectrum from NV2, using two different lengths of spin-echo sequences $\tau = 1 \, \mu$s and $\tau = 6 \, \mu$s respectively. The pulsewidth for the target spin is 220 ns, and the $\pi$-pulse length is 20 ns. The red curve in top plot is a fit to the data using the same sinc-squared function used in \figref{fig:4}, while the red curve in the bottom panel is a Lorentzian fit.}
\label{fig:5}
\end{figure}

Similar data can be obtained from other NVs, for instance DEER measurements with NV2 and NV3 are shown in \figref{fig:5} and \figref{fig:7} respectively, and displays several resonances.  NV2 is likely to be a native NV center (due to $^{14}$N hyperfine splitting observed in ODMR data), and has a longer coherence time $T_2 \sim 6 \,\mu$s, while NV3 is implanted and has similar $T_2$ as NV1. In \figref{fig:5}(a) the three resonances at 486 MHz, 811 MHz, and 1104 MHz with $\sim 300$ MHz splitting are likely due to the hyperfine structure of Cu$^{2+}$ as we explain below in Section~\ref{sec:simulation}. Another interesting feature is the difference in spectral linewidths and contrast in \figref{fig:5}(b) top and bottom plots, when we used two different spin-echo sequence lengths, which we also discuss below in Section~\ref{sec:simulation}. 

\subsection{DEER Rabi oscillations}
\begin{figure}[htb]
\includegraphics[width=0.7\textwidth]{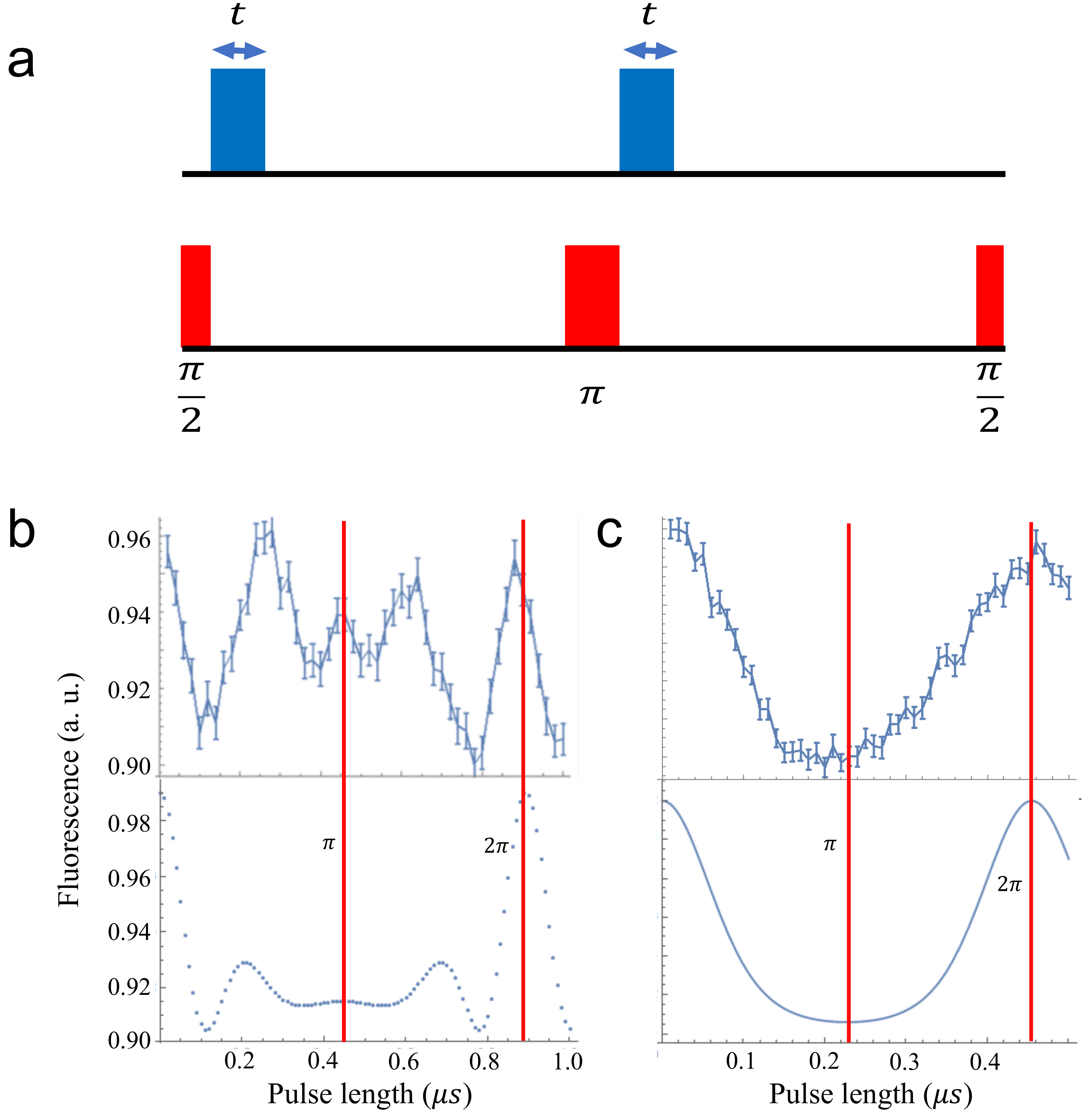}
\caption{(a) Experimental sequence for the DEER Rabi experiment. The frequency of the target spin driving pulse is fixed on DEER resonance, while the pulse length is scanned.  (b) (top panel)  Experimental DEER Rabi  data from NV1, the pulse width and frequency for NV1 is the same as in \figref{fig:4}, the frequency for the drive pulse on the target spin is 495 MHz. (bottom panel) Theory for DEER Rabi from a single target spin using $\gamma_e \tau \lambda (\hat{e}_B \cdot \hat{e}_{i}) = 5.8$ and $\Omega = 1.12$ MHz. The red lines mark the expected pulse lengths of the $\pi$ and $2 \pi$ pulses.(c) Experimental DEER Rabi  data from NV2, the pulse width and frequency for NV2 is the same as in \figref{fig:5}, the frequency for the drive pulse on the target spin is 486.4 MHz. (bottom panel) Theory for DEER Rabi from an ensemble of electron spins using the prefactors $ \tfrac{2n \bar{c}^2}{3} = 2.5$ and $\Omega = 2.2$~MHz . The red lines mark the expected pulse lengths of the $\pi$ and $2 \pi$ pulses.}
\label{fig:6}
\end{figure}

Our next set of experiments seeks to explore the strength of the coupling between NV and target spins. We carried out the DEER Rabi experiment with the pulse sequence shown in \figref{fig:6}(a). The frequency of the drive pulse is fixed at the DEER resonance observed in the earlier experiment, and the length of the drive pulse is scanned.  We used the following optimization procedure to obtain high quality data for both the DEER frequency spectrum and Rabi oscillations: (i) Perform DEER frequency scan with driving power selected at $\sim 0$ dBm and small pulse lengths, in frequency ranges where a DEER dip is expected. Considering our small linewidth, each step of the frequency scan is $\sim 2$ MHz at most. (ii) Keep averaging until fluctuation in spectrum becomes smooth and repeatable dip shows up. If no dips are discovered in the signal, jump to another frequency range, or change to another pulse length and try again. (iii) Use the estimated center value of the dip as the fixed driving frequency, and perform DEER Rabi experiment to find the proper $\pi$-pulse length. (iv) Perform DEER frequency scan to get a better spectrum with high contrast.

The results are shown in \figref{fig:6}(b) for NV1. As described in Section~\ref{sec:singleDEER}, the signal for the situation of a single target spin can be calculated, and by varying the pre-factor and the microwave parameters, the theoretical prediction closely matches the experimental data. From the data and our model, we obtain the expected values of the $\pi$ and $2 \pi$ pulse lengths as shown by the red vertical lines. The observation of oscillations within one $\pi$ period, and the good fit to our model with a single target spin, strengthens our interpretation that this signal is due to a single Cu$^{2+}$ electron spin. By contrast, the data with NV2, which is expected to be deeper due to its longer coherence time ($T_2 \sim 6 \,\mu$s), shows no oscillations, and is consistent with sensing of a small ensemble of electron spins. Our model for ensemble DEER gives a good fit to the data as shown in the bottom panel of \figref{fig:6}(c). However, another possibility that we cannot totally rule out is that the data for NV2 arises from a single target electron spin with coupling constant $c \sim 1$ as shown in \figref{fig:2}(b).

\begin{figure}[htb]
\includegraphics[width=\textwidth]{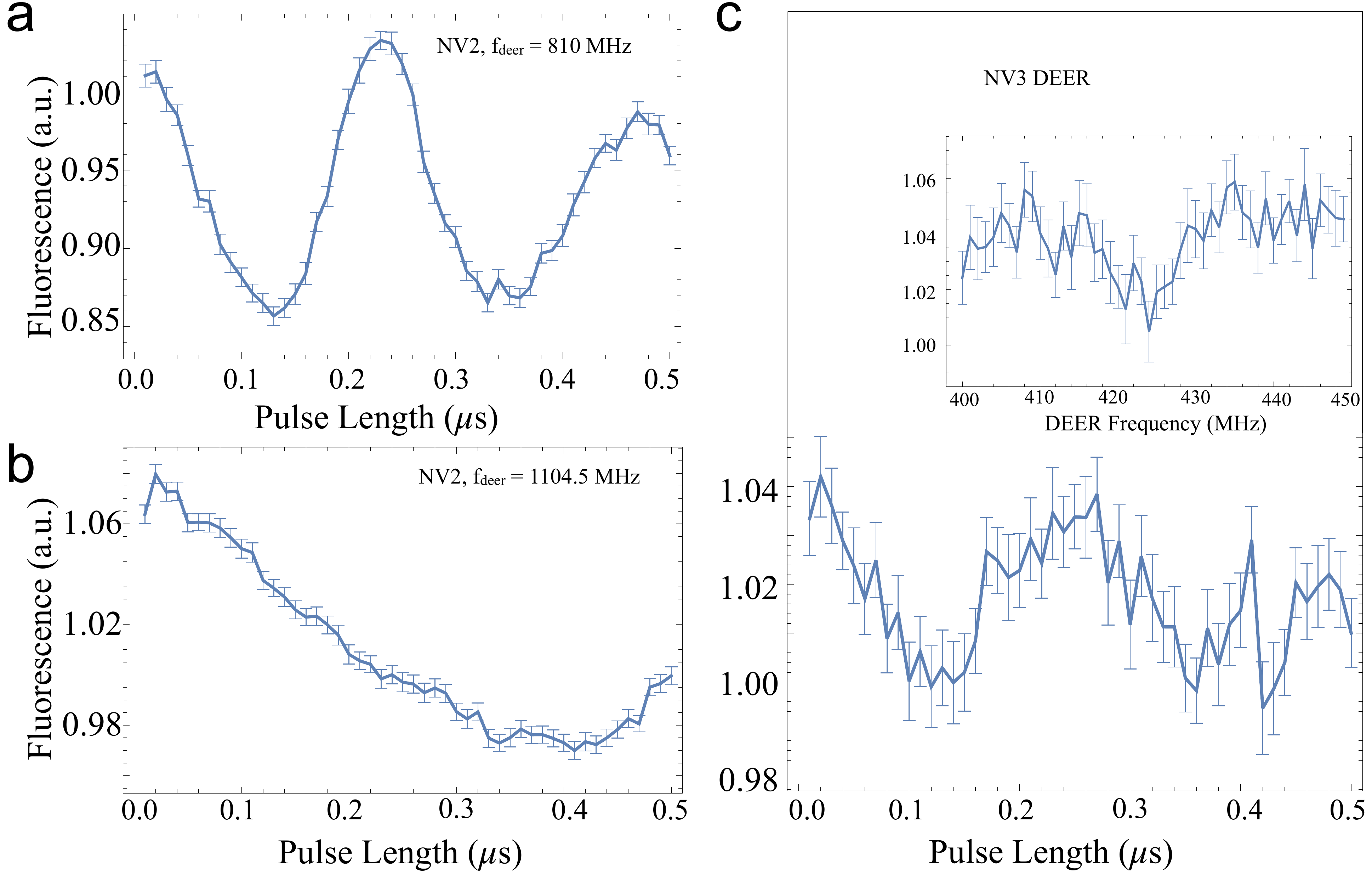}
\caption{(a) DEER Rabi data for NV2 with the frequency for the drive pulse on target spin at 810 MHz. (b) DEER Rabi data for NV2 with the frequency for the drive pulse on target spin at 1104.5 MHz. (c) DEER Rabi data for NV3 with the drive pulse frequency at 423.5 MHz. (inset) Experimental DEER spectrum from NV3.}
\label{fig:7}
\end{figure}

In \figref{fig:7}(a), we show the DEER Rabi data for NV2 at the other two resonance peaks $f_{deer} = 810 \text{ and } 1104.5$ MHz. For completeness \figref{fig:7}(b) also show the DEER Rabi data from another sensor NV (NV3). The low signal to noise for NV3 is typical for many of the NVs we observed in our sample, possibly due to smaller coupling strength of target electron spins near the sensor NVs. This possibility is further discussed in Section~\ref{sec:analysis}. 

\section{Analysis and Discussion}
\label{sec:analysis}
The observed resonances could potentially arise from intrinsic defects in diamond such as the commonly occurring P1 center. The P1 center Hamiltonian is given by~\cite{Loubser:1978aa,Abeywardana:2016aa},
\begin{equation}
    H_{P1} = \mu_B \vec{B} \cdot \overset\leftrightarrow{g}_{P1} \cdot \vec{S} + \vec{I}_N \cdot \overset\leftrightarrow{A} \cdot \vec{S} - g_n \mu_n \vec{B} \cdot \vec{I}_N - P_z (I_{N,z})^2
\end{equation}
where $\vec{S}$ is the electron spin operator for the P1 center with spin 1/2, $\vec{I}_N$ is the $^{14}$N nuclear spin operator with spin 1, $g_{P1}$ is the axially symmetric g-tensor of the P1 center $g_{x} = g_{y} = -2.0024$, $g_z = -2.0025$, $\mu_B$ is the Bohr magneton, the hyperfine interaction tensor $\vec{\boldsymbol{A}} = \text{diag(82, 82, 114)}$ MHz, $g_n = 0.403$ is the nuclear g-factor, $\mu_n$ is the nuclear magneton,  and $P_z = -5.6$ MHz is the quadrupole field strength. Solving this Hamiltonian with our applied magnetic field $B \approx (114,0,163)$ Gauss where the z-axis is along the [111] crystallographic direction of the NV and P1 center, we obtain the expected positions of the peaks as $\approx 79, 188, 231$ MHz. The observed resonances in our experiment differ from these values significantly both in position and in the hyperfine splitting between the peaks. 
 
 Similarly, the DEER signals arising from paramagnetic dangling bonds on the diamond surface~\cite{Sushkov2014a} would have resonance frequency $\sim 560$ MHz.  Other intrinsic defects arising from nitrogen centers such as the N1, W7, and P2 centers all have similar g-factors and hyperfine splittings as the P1 center with variations $\sim 10 \%$ (see Ref.~\cite{Loubser:1978aa} for a comprehensive review). We cannot rule out of course all the other possible intrinsic defects from transition metal ions which could give rise to similar signals as we observed, but it seems unlikely to appear in three different NV locations with slightly different resonance frequencies. For instance, NV2 and NV3 data was taken at nearly the same magnetic fields, but the DEER resonance frequencies for the lowest peaks differ by nearly 60 MHz.  As discussed below in Section~\ref{sec:simulation}, variations in the angle of the principal axes for the Cu$^{2+}$ ion with respect to the magnetic field could account for these variations in the observed DEER spectra. Interestingly, the DEER signals from P1 centers as well as the dangling bonds of surface states on diamond also tend to be broader than the resonances we have observed here~\cite{Sushkov2014a,Stepanov:2016aa,Fortman:2019aa}.

The observed DEER spectrum is not consistent with a powder spectrum of Cu$^{2+}$ ions commonly seen in high-field EPR studies~\cite{Yang2015:bb,Cunningham:2015aa,Ji:2014aa}. The linewidth we observed is also much narrower than the DEER signals from nitroxyl group or quantum sensing work carried out in Refs.~\cite{Mamin2012:bb,Shi:2015aa}.  If the signals were caused by clusters of metal ion complexes coordinated by chloride or other ligand molecules on the surface, we would expect very strong dipolar interactions between the spins which would cause significant broadening of the linewidth. In a cluster, the broadening due to dipole-dipole interaction arises from random orientation of the spins and the anisotropy of the interaction. In single crystals, however, the spins are well-aligned and at low concentrations the spectrum is consequently narrow, while at higher concentrations dipolar and exchange interactions can lead to a broadened lines with some narrow peaks~\cite{santana_single_2005,neuman_single_2010,neuman_single_2012}. We interpret the narrow and difficult to measure DEER signals as possibly arising from single Cu ions or small ensembles of Cu ions separated by larger ($\approx 2 - 10$ nm) distances trapped in the polymer matrix.  Small nano-crystals of CuCl$_2$ trapped in the polymer are an alternate explanation but we have no independent confirmation of such crystals. We hypothesize that the sample exists in a heterogenous state with mixture of clusters of metal ions, single or small ensembles of separated metal ions, and possibly nanocrystals distributed randomly and trapped by the polymer matrix. The clusters of metal ions would have large dipolar broadening and be difficult to detect due to the lower signal to noise. The few single ions (or possibly nano-crystals) that are close to the sensor NV would then give rise to the narrow DEER signals we observe. In the sections below, we analyze the Hamiltonian of the Cu$^{2+}$ spin, compare the simulations to the experimental results from DEER spectra, and estimate the approximate sensing volume.

\subsection{Density of target copper spins}
To estimate the number of target copper spins in our sensing volume, we assume the CuCl$_2$ molecules are uniformly distributed in the poly-L-lysine residuals. We can estimate the volume of the poly-L-lysine deposited as $4.4 \times 10^{-4} \, \text{mm}^3$. Similarly, based on the volume of the CuCl$_2$ crystals which is less than one-tenth of the volume of the poly-L-lysine, we can estimate that we have $\sim 0.5$~nmol of Cu$^{2+}$ ions , and therefore the average number of Cu$^{2+}$ ions per unit volume as $\sim 0.6 \, \text{spins/}\text{(nm)}^3$ , i.e. the volume per spin is $\sim 1.6 \, \text{nm}^3$. The dipolar interaction between electron spins in copper at that distance is $\sim 40$ MHz which is small compared to the hyperfine interaction, but non-negligible. Given the area of the sample as $\sim 4 \, \text{mm}^2$, we can also estimate that the thickness of the poly-L-lysine residuals as $\ll 100$~nm in the working area of the sample where we observe the signals. 

\subsection{Sensing volume}
\begin{figure}[htb]
\includegraphics[width=\textwidth]{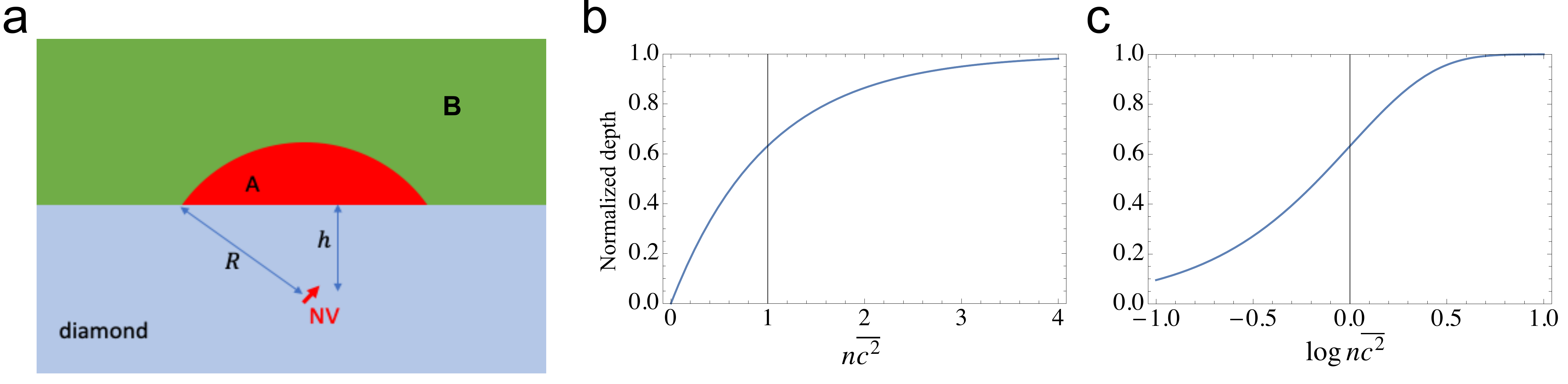}
\caption{(a) Schematic drawing of the sensing volume: the spherical cap region (marked A) with radius $R$ shown in red, is our sensing volume. The volume B, which is the non-detectable region, refers to the region where spins dont contribute with sufficient intensity to be detectable. (b), (c) Normalized depth of the DEER spectral dips as a function of the parameter $n \bar{c}^2$, shown both in linear and log scale on the x-axis respectively. Our threshold of detectability is chosen as $n \bar{c}^2 = 1$, marked with black vertical lines in the plot, and any DEER signal with $n \bar{c}^2 < 1$ is treated as undetectable.}
\label{fig:8}
\end{figure}

As discussed in Section~\ref{sec:ensembleDEER}, all electron spins from Cu$^{2+}$ ions in the sensing volume contribute to the DEER signal. Setting up the boundary of the sensing volume then depends on our threshold of detectable signal, and the shape of the boundary we choose. As shown in \figref{fig:2}, the parameter $n \bar{c}^2$ determines the magnitude of the DEER signal, and we can choose our threshold of detectability based on this factor. The normalized DEER signal is plotted as a function of $ n \bar{c}^2$ in \figref{fig:8}(a), and we see that the signal drops below 70\% of the maximum at $n \bar{c}^2 = 1$. Hence we choose this as our threshold.

Based on the theory of DEER we discussed in Section~\ref{sec:singleDEER} and Section~\ref{sec:ensembleDEER} , all electron spins contribute to the parameter $n \bar{c}^2 = \sum_k c_k^2$ as,
\begin{equation}
\lvert c_k \rvert = \frac{\mu_0 \gamma_e \hbar^2 \tau}{8 \pi} . \frac{1}{r_k^3} . \lvert \vec{e}_B \cdot \hat{e}_{i,k} \rvert
\end{equation}
The last term varies due to the different directions for different Cu$^{2+}$ but typically will be of order unity. We therefore approximate that,
\begin{equation}
\lvert c_k \rvert = c(r)  = \kappa . \frac{1}{r^3}
\end{equation}
where $\kappa = \frac{\mu_0 \gamma_e \hbar^2 \tau}{8 \pi} \approx (9.9 \, \text{nm})^3$ is a constant. The contribution to the parameter $n \bar{c}^2$ from different electrons can therefore be binned into spherical shells with different distances to the NV center. Thus, we can define our sensing volumes as spherical regions shown in \figref{fig:8}(B). This simplified spherical region model of sensing volume is also used in Ref.~\cite{Staudacher2013}. We use the same division of our sensing volume as in that work, such that 70\% of the total signal is contributed from the sensing volume of the spherical (red) region shown in \figref{fig:8}(b) while the remaining 30\% of the signal comes from the rest of the (green) region, and contributes a signal that is below our detectable threshold. Using this method, we can even estimate the depth of our NV center $h$ from the measured signal. We also apply the following rules to estimate our sensing volume: (i) Assuming the maximum number density of electron spins, the contribution from the sensing volume is detectable above our threshold. (ii) Even with maximum number density of electrons in the non-detectable region, the contribution from the volume is below the threshold.

From these rules, we infer that the depth of the NV center $h$ cannot be greater than 170 nm, in order to sense a detectable signal from the target electron spins. In fact, since our DEER Rabi experiment for NV2 best fits to $n \bar{c}^2 \sim 5$, the depth of NV2 is precluded from being above 100 nm. Further, assuming the depth of NV2 $ h < 100$~nm, the maximum radius of the sensing volume is lesser than $\sim 250$~nm. Anything outside this radius will definitely not be detectable. Using all this information, and taking our data into consideration, we can arrive at a best estimate for the depth of NV2 as $h \sim 70$~nm and a sensing volume radius of $ 240 \text{ nm}$. NV1 and NV3 are likely much closer to the diamond surface than NV2 as they were implanted NVs but without more careful measurements of the resonances as a function of magnetic field and angle we cannot determine the exact depth more accurately.
 
 
\subsection{Simulation of the EPR spectrum}
\label{sec:simulation}
The Hamiltonian of a single Cu$^{2+}$ ion's unpaired electron spin in its $3d_{x^2 - y^2}$ orbital can be written as,
\begin{equation}
H =  \mu_B \vec{B} \cdot \overset\leftrightarrow{g} \cdot \vec{S} + \vec{I} \cdot \overset\leftrightarrow{A} \cdot \vec{S} 
\end{equation}
where $ \overset\leftrightarrow{g}$ is the anisotropic g-tensor, $ \overset\leftrightarrow{A}$ is the hyperfine interaction. The anisotropy in the g-tensor is typical in transition metal ions due to the spin-orbit coupling. For Cu$^{2+}$ ion, the g-tensor is axially symmetric ($g_x = g_y < g_z$). The second term is the hyperfine interaction due to the copper nuclear spin with $ I = 3/2$. The nuclear Zeeman and quadrupole interactions are neglected. The diagonal forms of the g-tensor and hyperfine interaction tensor along the principal axes are $\overset\leftrightarrow{g} = \text{diag}(g_{\perp}, g_{\perp},g_{\parallel})$ and $\overset\leftrightarrow{A} = \text{diag}(A_{\perp}, A_{\perp},A_{\parallel})$ where $g_{\perp} = -2.0835$, $g_{\parallel} = -2.415$, $ A_{\perp} = 30 \, \text{MHz}$, $ A_{\parallel} = 339 \, \text{MHz}$. For Cu$^{2+}$ ions in a single molecule or single crystal, the principal axes of $ \overset\leftrightarrow{g}$ and $ \overset\leftrightarrow{A}$ are the same in both frames since they are set up by the same 3d orbital bond orientation. We neglected dipolar interactions between the Cu electron spins. We used EasySpin to carry out a numerical simulation of the EPR spectrum of CuCl$_2$~\cite{Stoll:2006aa}. By adjusting the magnetic field $B$ and the angle $\theta$ between the field and the principal axis, we can minimize the sum of the squared distance between the peaks from the simulation and the experimentally observed resonances in the data ($\chi^2$ minimization). 

\begin{figure}[htb]
\includegraphics[width=0.9\textwidth]{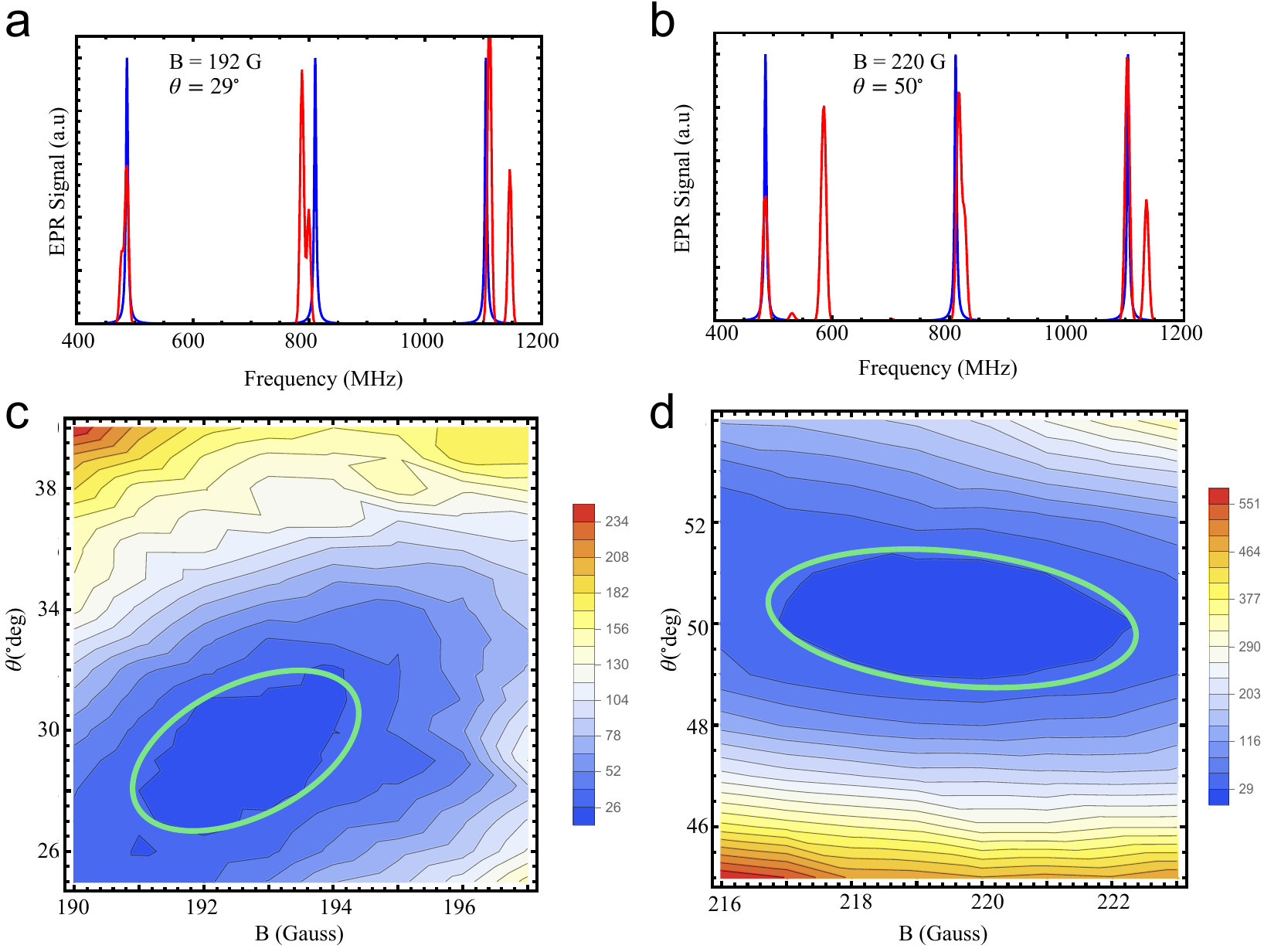}
\caption{(a),(b) Comparison of CuCl$_2$ EPR spectrum simulation and DEER experimental data for best fit parameter values of $(B,\theta)$ from panels (c),(d). The simulation for CuCl$_2$, carried out with EasySpin, are shown in red, and the blue peaks mark the observed position of resonances from our DEER spectrum. (c),(d) Goodness of fit ($\chi^2$) plotted as a function of $(B, \theta)$ where $\theta$ is the angle between the magnetic field and the principal axes of the Cu$^{2+}$ ion as explained in the text. Two different parameter sets for $(B,\theta)$ (denoted by green ovals) both represent reasonably good fits to our observed spectrum.}
\label{fig:9}
\end{figure}

We found two different parameter sets $(B, \theta)$ which gives reasonably good fits to the DEER spectrum. \figref{fig:9}(a) and (b) shows the comparison between the numerical simulation and our experimental DEER spectrum. \figref{fig:9}(c) and (d) shows the goodness of fit parameter $\chi^2$ contour plots as a function of $(B, \theta)$. The green highlighted contours in \figref{fig:9}(c) and (d) represents the best goodness of fit, from which we can also extract the uncertainty in the fit parameters which makes the goodness of fit increase by $\Delta \chi^2 = 1$. For these two local minima, we obtain $ B = 192 \pm 1 \, \text{G},\theta = 29 \pm 1^\circ$ and $ B = 220 \pm 1 \, \text{G},\theta = 50 \pm 1^\circ$.  Further, the two theoretical spectra clearly differ, with a strong resonance expected near $\approx 580$~MHz for the $ (B = 220 \, \text{G},\theta = 50^\circ)$ parameter set in \figref{fig:9}(b) that we did not observe. As explained in Section~\ref{sec:expdeerspectrum}, finding narrow peaks over a broad frequency range requires careful tuning up of the drive pulse parameters, so although we did not observe this peak we cannot definitively rule out the second parameter set. However, given our magnetic field values $B \approx (114, 0, 163)$ Gauss with a magnitude $| B | \approx 200$ Gauss, we conclude that the first value is much more likely. We were unable to scan for the smaller peak near $\approx 1150$~MHz that is present in both parameter sets due to experimental limitations.  Further experimental scans searching for these resonances after careful simulations will likely resolve the discrepancies, but is outside the scope of this work. 

Another interesting experimental detail is noted in \figref{fig:5}. In the top plot, when $\tau =1 \, \mu$s is less than the collapse time of the NV spin-echo sequence, the DEER spectrum is much narrower and has a higher contrast, and also displays oscillations. Whereas in the bottom plot, when we fix the time $\tau$ at the C$^{13}$ revival time of $6 \, \mu$s, the contrast is decreased and the oscillations seem to disappear. One possible reason could be spectral diffusion of the target Cu(II)-ions over the longer time scales of the experiment in the second situation, but we cannot rule out other reasons such as mechanical or thermal shifts. Such broadening and spectral diffusion would be interesting to examine as it may give us more information about the environment of the molecules being studied. Due to the tunability of the spin-echo filter frequency by varying the $\tau$, we hope to study this in more detail in future experiments. 

\section{Conclusions}
The DEER signals from Cu$^{2+}$ observed in this work demonstrate the nanoscale sensing of external electron spins bound to a transition-metal ion under room temperature, low magnetic field conditions. We observe surprisingly narrow resonances which imply that Cu-labeled molecules could be a viable candidate for distance measurements with high precision in biologically relevant molecules and nanoscale materials. The observation of oscillations within a collapse-revival time of the DEER Rabi signal is consistent with a single electron spin sensitivity. Further investigation into why the resonances are so narrow, given our power broadened pulse lengths, and measurements of $T_1$ of the target spins remains to be carried out. The dependence of the DEER spectral width on the spin-echo sequence length or characteristic filter frequency, would also be interesting to examine. New samples with NV centers implanted closer to the surface, application of multi-pulse dynamical decoupling sequences~\cite{Staudacher2013,Kolkowitz2012distantnucspin,Mamin2013}, and lower concentration of the target molecules will also be helpful in better understanding of the data and the models. 

\section{Acknowledgments}
This work was primarily funded by NSF EFRI ACQUIRE Grant No. 1741656. K.Zhang was also partially supported by Guangdong Special Support Project (2019BT02X030). The authors wish to thank Bradley Slezak and Brian D'Urso for help with annealing diamond samples, Fedor Jelezko and Priya Balasubramanian for helpful discussion about the DEER signals.


\begin{thebibliography}{50}%
\makeatletter
\providecommand \@ifxundefined [1]{%
 \@ifx{#1\undefined}
}%
\providecommand \@ifnum [1]{%
 \ifnum #1\expandafter \@firstoftwo
 \else \expandafter \@secondoftwo
 \fi
}%
\providecommand \@ifx [1]{%
 \ifx #1\expandafter \@firstoftwo
 \else \expandafter \@secondoftwo
 \fi
}%
\providecommand \natexlab [1]{#1}%
\providecommand \enquote  [1]{``#1''}%
\providecommand \bibnamefont  [1]{#1}%
\providecommand \bibfnamefont [1]{#1}%
\providecommand \citenamefont [1]{#1}%
\providecommand \href@noop [0]{\@secondoftwo}%
\providecommand \href [0]{\begingroup \@sanitize@url \@href}%
\providecommand \@href[1]{\@@startlink{#1}\@@href}%
\providecommand \@@href[1]{\endgroup#1\@@endlink}%
\providecommand \@sanitize@url [0]{\catcode `\\12\catcode `\$12\catcode
  `\&12\catcode `\#12\catcode `\^12\catcode `\_12\catcode `\%12\relax}%
\providecommand \@@startlink[1]{}%
\providecommand \@@endlink[0]{}%
\providecommand \url  [0]{\begingroup\@sanitize@url \@url }%
\providecommand \@url [1]{\endgroup\@href {#1}{\urlprefix }}%
\providecommand \urlprefix  [0]{URL }%
\providecommand \Eprint [0]{\href }%
\providecommand \doibase [0]{http://dx.doi.org/}%
\providecommand \selectlanguage [0]{\@gobble}%
\providecommand \bibinfo  [0]{\@secondoftwo}%
\providecommand \bibfield  [0]{\@secondoftwo}%
\providecommand \translation [1]{[#1]}%
\providecommand \BibitemOpen [0]{}%
\providecommand \bibitemStop [0]{}%
\providecommand \bibitemNoStop [0]{.\EOS\space}%
\providecommand \EOS [0]{\spacefactor3000\relax}%
\providecommand \BibitemShut  [1]{\csname bibitem#1\endcsname}%
\let\auto@bib@innerbib\@empty
\bibitem [{\citenamefont {Hubbell}\ \emph {et~al.}(2013)\citenamefont
  {Hubbell}, \citenamefont {Lopez}, \citenamefont {Altenbach},\ and\
  \citenamefont {Yang}}]{Hubbell:2013aa}%
  \BibitemOpen
  \bibfield  {author} {\bibinfo {author} {\bibfnamefont {W.~L.}\ \bibnamefont
  {Hubbell}}, \bibinfo {author} {\bibfnamefont {C.~J.}\ \bibnamefont {Lopez}},
  \bibinfo {author} {\bibfnamefont {C.}~\bibnamefont {Altenbach}}, \ and\
  \bibinfo {author} {\bibfnamefont {Z.}~\bibnamefont {Yang}},\ }\href@noop {}
  {\bibfield  {journal} {\bibinfo  {journal} {Curr. Opin. Struct. Biol.}\
  }\textbf {\bibinfo {volume} {23}},\ \bibinfo {pages} {725} (\bibinfo {year}
  {2013})}\BibitemShut {NoStop}%
\bibitem [{\citenamefont {Ubbink}\ \emph {et~al.}(2002)\citenamefont {Ubbink},
  \citenamefont {Worrall}, \citenamefont {Canters}, \citenamefont {Groenen},\
  and\ \citenamefont {Huber}}]{Ubbink:2002aa}%
  \BibitemOpen
  \bibfield  {author} {\bibinfo {author} {\bibfnamefont {M.}~\bibnamefont
  {Ubbink}}, \bibinfo {author} {\bibfnamefont {J.~A.~R.}\ \bibnamefont
  {Worrall}}, \bibinfo {author} {\bibfnamefont {G.~W.}\ \bibnamefont
  {Canters}}, \bibinfo {author} {\bibfnamefont {E.~J.~J.}\ \bibnamefont
  {Groenen}}, \ and\ \bibinfo {author} {\bibfnamefont {M.}~\bibnamefont
  {Huber}},\ }\href@noop {} {\bibfield  {journal} {\bibinfo  {journal} {Annual
  Review of Biophysics and Biomolecular Structure}\ }\textbf {\bibinfo {volume}
  {31}},\ \bibinfo {pages} {393} (\bibinfo {year} {2002})}\BibitemShut
  {NoStop}%
\bibitem [{\citenamefont {Feher}(1959)}]{Feher:1959aa}%
  \BibitemOpen
  \bibfield  {author} {\bibinfo {author} {\bibfnamefont {G.}~\bibnamefont
  {Feher}},\ }\href@noop {} {\bibfield  {journal} {\bibinfo  {journal} {Phys.
  Rev.}\ }\textbf {\bibinfo {volume} {114}},\ \bibinfo {pages} {1219} (\bibinfo
  {year} {1959})}\BibitemShut {NoStop}%
\bibitem [{\citenamefont {Watkins}(1998)}]{Watkins:1998aa}%
  \BibitemOpen
  \bibfield  {author} {\bibinfo {author} {\bibfnamefont {G.~D.}\ \bibnamefont
  {Watkins}},\ }\href@noop {} {\bibfield  {journal} {\bibinfo  {journal}
  {Semicond. Semimetals}\ }\textbf {\bibinfo {volume} {51A}},\ \bibinfo {pages}
  {1} (\bibinfo {year} {1998})}\BibitemShut {NoStop}%
\bibitem [{\citenamefont {Loubser}\ and\ \citenamefont {van
  Wyk}(1978)}]{Loubser:1978aa}%
  \BibitemOpen
  \bibfield  {author} {\bibinfo {author} {\bibfnamefont {J.~H.~N.}\
  \bibnamefont {Loubser}}\ and\ \bibinfo {author} {\bibfnamefont {J.~A.}\
  \bibnamefont {van Wyk}},\ }\href@noop {} {\bibfield  {journal} {\bibinfo
  {journal} {Rep. Prog. Phys.}\ }\textbf {\bibinfo {volume} {41}},\ \bibinfo
  {pages} {1201} (\bibinfo {year} {1978})}\BibitemShut {NoStop}%
\bibitem [{\citenamefont {Davydov}\ \emph {et~al.}(2009)\citenamefont
  {Davydov}, \citenamefont {Sudhamsu}, \citenamefont {Lees}, \citenamefont
  {Crane},\ and\ \citenamefont {Hoffman}}]{Davydon:2009aa}%
  \BibitemOpen
  \bibfield  {author} {\bibinfo {author} {\bibfnamefont {R.}~\bibnamefont
  {Davydov}}, \bibinfo {author} {\bibfnamefont {J.}~\bibnamefont {Sudhamsu}},
  \bibinfo {author} {\bibfnamefont {N.~S.}\ \bibnamefont {Lees}}, \bibinfo
  {author} {\bibfnamefont {B.~R.}\ \bibnamefont {Crane}}, \ and\ \bibinfo
  {author} {\bibfnamefont {B.~M.}\ \bibnamefont {Hoffman}},\ }\href@noop {}
  {\bibfield  {journal} {\bibinfo  {journal} {J. Am. Chem. Soc.}\ }\textbf
  {\bibinfo {volume} {131}},\ \bibinfo {pages} {14493} (\bibinfo {year}
  {2009})}\BibitemShut {NoStop}%
\bibitem [{\citenamefont {Niklas}\ and\ \citenamefont
  {Poluektov}(2017)}]{Niklas:2017aa}%
  \BibitemOpen
  \bibfield  {author} {\bibinfo {author} {\bibfnamefont {J.}~\bibnamefont
  {Niklas}}\ and\ \bibinfo {author} {\bibfnamefont {O.~G.}\ \bibnamefont
  {Poluektov}},\ }\href {\doibase 10.1002/aenm.201770053} {\bibfield  {journal}
  {\bibinfo  {journal} {Adv. Energy Mater.}\ } (\bibinfo {year} {2017}),\
  10.1002/aenm.201770053}\BibitemShut {NoStop}%
\bibitem [{\citenamefont {Shaakov}\ \emph {et~al.}(2003)\citenamefont
  {Shaakov}, \citenamefont {Galili}, \citenamefont {Stavitski}, \citenamefont
  {Levanon}, \citenamefont {Lukas},\ and\ \citenamefont
  {Wasielewski}}]{Shaakov:2003aa}%
  \BibitemOpen
  \bibfield  {author} {\bibinfo {author} {\bibfnamefont {S.}~\bibnamefont
  {Shaakov}}, \bibinfo {author} {\bibfnamefont {T.}~\bibnamefont {Galili}},
  \bibinfo {author} {\bibfnamefont {E.}~\bibnamefont {Stavitski}}, \bibinfo
  {author} {\bibfnamefont {H.}~\bibnamefont {Levanon}}, \bibinfo {author}
  {\bibfnamefont {A.}~\bibnamefont {Lukas}}, \ and\ \bibinfo {author}
  {\bibfnamefont {M.~R.}\ \bibnamefont {Wasielewski}},\ }\href@noop {}
  {\bibfield  {journal} {\bibinfo  {journal} {J. Am. Chem. Soc.}\ }\textbf
  {\bibinfo {volume} {125}},\ \bibinfo {pages} {6563} (\bibinfo {year}
  {2003})}\BibitemShut {NoStop}%
\bibitem [{\citenamefont {Britt}\ \emph {et~al.}(2000)\citenamefont {Britt},
  \citenamefont {Peloquin},\ and\ \citenamefont {Campbell}}]{Britt:2000aa}%
  \BibitemOpen
  \bibfield  {author} {\bibinfo {author} {\bibfnamefont {R.~D.}\ \bibnamefont
  {Britt}}, \bibinfo {author} {\bibfnamefont {J.~M.}\ \bibnamefont {Peloquin}},
  \ and\ \bibinfo {author} {\bibfnamefont {K.~A.}\ \bibnamefont {Campbell}},\
  }\href@noop {} {\bibfield  {journal} {\bibinfo  {journal} {Annu. Rev.
  Biophys.}\ }\textbf {\bibinfo {volume} {29}},\ \bibinfo {pages} {464}
  (\bibinfo {year} {2000})}\BibitemShut {NoStop}%
\bibitem [{\citenamefont {J.M.Taylor}\ \emph {et~al.}(2008)\citenamefont
  {J.M.Taylor}, \citenamefont {Cappellaro}, \citenamefont {Childress},
  \citenamefont {Jiang}, \citenamefont {D.Budker}, \citenamefont {P.R.Hemmer},
  \citenamefont {A.Yacoby}, \citenamefont {Walsworth},\ and\ \citenamefont
  {Lukin}}]{Taylor08}%
  \BibitemOpen
  \bibfield  {author} {\bibinfo {author} {\bibnamefont {J.M.Taylor}}, \bibinfo
  {author} {\bibfnamefont {P.}~\bibnamefont {Cappellaro}}, \bibinfo {author}
  {\bibfnamefont {L.}~\bibnamefont {Childress}}, \bibinfo {author}
  {\bibfnamefont {L.}~\bibnamefont {Jiang}}, \bibinfo {author} {\bibnamefont
  {D.Budker}}, \bibinfo {author} {\bibnamefont {P.R.Hemmer}}, \bibinfo {author}
  {\bibnamefont {A.Yacoby}}, \bibinfo {author} {\bibfnamefont {R.}~\bibnamefont
  {Walsworth}}, \ and\ \bibinfo {author} {\bibfnamefont {M.~D.}\ \bibnamefont
  {Lukin}},\ }\href@noop {} {\bibfield  {journal} {\bibinfo  {journal} {Nature
  Physics}\ }\textbf {\bibinfo {volume} {4}},\ \bibinfo {pages} {810} (\bibinfo
  {year} {2008})}\BibitemShut {NoStop}%
\bibitem [{\citenamefont {Maze}\ \emph {et~al.}(2008)\citenamefont {Maze},
  \citenamefont {Stanwix}, \citenamefont {Hodges}, \citenamefont {Hong},
  \citenamefont {Taylor}, \citenamefont {Cappellaro}, \citenamefont {Jiang},
  \citenamefont {Togan}, \citenamefont {Dutt}, \citenamefont {Zibrov},
  \citenamefont {Yacoby}, \citenamefont {Walsworth},\ and\ \citenamefont
  {Lukin}}]{Maze08}%
  \BibitemOpen
  \bibfield  {author} {\bibinfo {author} {\bibfnamefont {J.~R.}\ \bibnamefont
  {Maze}}, \bibinfo {author} {\bibfnamefont {P.~L.}\ \bibnamefont {Stanwix}},
  \bibinfo {author} {\bibfnamefont {J.~S.}\ \bibnamefont {Hodges}}, \bibinfo
  {author} {\bibfnamefont {S.}~\bibnamefont {Hong}}, \bibinfo {author}
  {\bibfnamefont {J.~M.}\ \bibnamefont {Taylor}}, \bibinfo {author}
  {\bibfnamefont {P.}~\bibnamefont {Cappellaro}}, \bibinfo {author}
  {\bibfnamefont {L.}~\bibnamefont {Jiang}}, \bibinfo {author} {\bibfnamefont
  {E.}~\bibnamefont {Togan}}, \bibinfo {author} {\bibfnamefont {M.~V.~G.}\
  \bibnamefont {Dutt}}, \bibinfo {author} {\bibfnamefont {A.~S.}\ \bibnamefont
  {Zibrov}}, \bibinfo {author} {\bibfnamefont {A.}~\bibnamefont {Yacoby}},
  \bibinfo {author} {\bibfnamefont {R.~L.}\ \bibnamefont {Walsworth}}, \ and\
  \bibinfo {author} {\bibfnamefont {M.~D.}\ \bibnamefont {Lukin}},\ }\href@noop
  {} {\bibfield  {journal} {\bibinfo  {journal} {Nature}\ }\textbf {\bibinfo
  {volume} {455}},\ \bibinfo {pages} {644} (\bibinfo {year}
  {2008})}\BibitemShut {NoStop}%
\bibitem [{\citenamefont {Balasubramanian}\ \emph {et~al.}(2008)\citenamefont
  {Balasubramanian}, \citenamefont {Chan}, \citenamefont {Kolesov},
  \citenamefont {Al-Hmoud}, \citenamefont {Tisler}, \citenamefont {Shin},
  \citenamefont {Kim}, \citenamefont {Wojcik}, \citenamefont {Hemmer},
  \citenamefont {Krueger}, \citenamefont {Hanke}, \citenamefont
  {Leitenstorfer}, \citenamefont {Bratschitsch}, \citenamefont {Jelezko},\ and\
  \citenamefont {Wrachtrup}}]{Balasub08}%
  \BibitemOpen
  \bibfield  {author} {\bibinfo {author} {\bibfnamefont {G.}~\bibnamefont
  {Balasubramanian}}, \bibinfo {author} {\bibfnamefont {I.~Y.}\ \bibnamefont
  {Chan}}, \bibinfo {author} {\bibfnamefont {R.}~\bibnamefont {Kolesov}},
  \bibinfo {author} {\bibfnamefont {M.}~\bibnamefont {Al-Hmoud}}, \bibinfo
  {author} {\bibfnamefont {J.}~\bibnamefont {Tisler}}, \bibinfo {author}
  {\bibfnamefont {C.}~\bibnamefont {Shin}}, \bibinfo {author} {\bibfnamefont
  {C.}~\bibnamefont {Kim}}, \bibinfo {author} {\bibfnamefont {A.}~\bibnamefont
  {Wojcik}}, \bibinfo {author} {\bibfnamefont {P.~R.}\ \bibnamefont {Hemmer}},
  \bibinfo {author} {\bibfnamefont {A.}~\bibnamefont {Krueger}}, \bibinfo
  {author} {\bibfnamefont {T.}~\bibnamefont {Hanke}}, \bibinfo {author}
  {\bibfnamefont {A.}~\bibnamefont {Leitenstorfer}}, \bibinfo {author}
  {\bibfnamefont {R.}~\bibnamefont {Bratschitsch}}, \bibinfo {author}
  {\bibfnamefont {F.}~\bibnamefont {Jelezko}}, \ and\ \bibinfo {author}
  {\bibfnamefont {J.}~\bibnamefont {Wrachtrup}},\ }\href@noop {} {\bibfield
  {journal} {\bibinfo  {journal} {Nature}\ }\textbf {\bibinfo {volume} {455}},\
  \bibinfo {pages} {648} (\bibinfo {year} {2008})}\BibitemShut {NoStop}%
\bibitem [{\citenamefont {Degen}\ \emph {et~al.}(2017)\citenamefont {Degen},
  \citenamefont {Reinhard},\ and\ \citenamefont {Cappellaro}}]{Degen2017}%
  \BibitemOpen
  \bibfield  {author} {\bibinfo {author} {\bibfnamefont {C.~L.}\ \bibnamefont
  {Degen}}, \bibinfo {author} {\bibfnamefont {F.}~\bibnamefont {Reinhard}}, \
  and\ \bibinfo {author} {\bibfnamefont {P.}~\bibnamefont {Cappellaro}},\
  }\href {\doibase 10.1103/RevModPhys.89.035002} {\bibfield  {journal}
  {\bibinfo  {journal} {Reviews of Modern Physics}\ }\textbf {\bibinfo {volume}
  {89}},\ \bibinfo {pages} {1} (\bibinfo {year} {2017})},\ \Eprint
  {http://arxiv.org/abs/1611.02427} {arXiv:1611.02427} \BibitemShut {NoStop}%
\bibitem [{\citenamefont {Barry}\ \emph {et~al.}(2020)\citenamefont {Barry},
  \citenamefont {Schloss}, \citenamefont {Bauch}, \citenamefont {Turner},
  \citenamefont {Hart}, \citenamefont {Pham},\ and\ \citenamefont
  {Walsworth}}]{Barry2020}%
  \BibitemOpen
  \bibfield  {author} {\bibinfo {author} {\bibfnamefont {J.~F.}\ \bibnamefont
  {Barry}}, \bibinfo {author} {\bibfnamefont {J.~M.}\ \bibnamefont {Schloss}},
  \bibinfo {author} {\bibfnamefont {E.}~\bibnamefont {Bauch}}, \bibinfo
  {author} {\bibfnamefont {M.~J.}\ \bibnamefont {Turner}}, \bibinfo {author}
  {\bibfnamefont {C.~A.}\ \bibnamefont {Hart}}, \bibinfo {author}
  {\bibfnamefont {L.~M.}\ \bibnamefont {Pham}}, \ and\ \bibinfo {author}
  {\bibfnamefont {R.~L.}\ \bibnamefont {Walsworth}},\ }\href {\doibase
  10.1103/RevModPhys.92.015004} {\bibfield  {journal} {\bibinfo  {journal}
  {Rev. Mod. Phys.}\ }\textbf {\bibinfo {volume} {92}},\ \bibinfo {pages}
  {015004} (\bibinfo {year} {2020})}\BibitemShut {NoStop}%
\bibitem [{\citenamefont {Casola}\ \emph {et~al.}(2018)\citenamefont {Casola},
  \citenamefont {{Van Der Sar}},\ and\ \citenamefont {Yacoby}}]{Casola2018}%
  \BibitemOpen
  \bibfield  {author} {\bibinfo {author} {\bibfnamefont {F.}~\bibnamefont
  {Casola}}, \bibinfo {author} {\bibfnamefont {T.}~\bibnamefont {{Van Der
  Sar}}}, \ and\ \bibinfo {author} {\bibfnamefont {A.}~\bibnamefont {Yacoby}},\
  }\href {\doibase 10.1038/natrevmats.2017.88} {\bibfield  {journal} {\bibinfo
  {journal} {Nature Reviews Materials}\ }\textbf {\bibinfo {volume} {3}}
  (\bibinfo {year} {2018}),\ 10.1038/natrevmats.2017.88},\ \Eprint
  {http://arxiv.org/abs/1804.08742} {arXiv:1804.08742} \BibitemShut {NoStop}%
\bibitem [{\citenamefont {Staudacher}\ \emph {et~al.}(2013)\citenamefont
  {Staudacher}, \citenamefont {Shi}, \citenamefont {Pezzagna}, \citenamefont
  {Meijer}, \citenamefont {Du}, \citenamefont {Meriles}, \citenamefont
  {Reinhard},\ and\ \citenamefont {Wrachtrup}}]{Staudacher2013}%
  \BibitemOpen
  \bibfield  {author} {\bibinfo {author} {\bibfnamefont {T.}~\bibnamefont
  {Staudacher}}, \bibinfo {author} {\bibfnamefont {F.}~\bibnamefont {Shi}},
  \bibinfo {author} {\bibfnamefont {S.}~\bibnamefont {Pezzagna}}, \bibinfo
  {author} {\bibfnamefont {J.}~\bibnamefont {Meijer}}, \bibinfo {author}
  {\bibfnamefont {J.}~\bibnamefont {Du}}, \bibinfo {author} {\bibfnamefont
  {C.~a.}\ \bibnamefont {Meriles}}, \bibinfo {author} {\bibfnamefont
  {F.}~\bibnamefont {Reinhard}}, \ and\ \bibinfo {author} {\bibfnamefont
  {J.}~\bibnamefont {Wrachtrup}},\ }\href {\doibase 10.1126/science.1231675}
  {\bibfield  {journal} {\bibinfo  {journal} {Science (New York, N.Y.)}\
  }\textbf {\bibinfo {volume} {339}},\ \bibinfo {pages} {561} (\bibinfo {year}
  {2013})},\ \Eprint {http://arxiv.org/abs/arXiv:1011.1669v3}
  {arXiv:arXiv:1011.1669v3} \BibitemShut {NoStop}%
\bibitem [{\citenamefont {Mamin}\ \emph {et~al.}(2013)\citenamefont {Mamin},
  \citenamefont {Kim}, \citenamefont {Sherwood}, \citenamefont {Rettner},
  \citenamefont {Ohno}, \citenamefont {Awschalom},\ and\ \citenamefont
  {Rugar}}]{Mamin2013}%
  \BibitemOpen
  \bibfield  {author} {\bibinfo {author} {\bibfnamefont {H.~J.}\ \bibnamefont
  {Mamin}}, \bibinfo {author} {\bibfnamefont {M.}~\bibnamefont {Kim}}, \bibinfo
  {author} {\bibfnamefont {M.~H.}\ \bibnamefont {Sherwood}}, \bibinfo {author}
  {\bibfnamefont {C.~T.}\ \bibnamefont {Rettner}}, \bibinfo {author}
  {\bibfnamefont {K.}~\bibnamefont {Ohno}}, \bibinfo {author} {\bibfnamefont
  {D.~D.}\ \bibnamefont {Awschalom}}, \ and\ \bibinfo {author} {\bibfnamefont
  {D.}~\bibnamefont {Rugar}},\ }\href {\doibase 10.1126/science.1231540}
  {\bibfield  {journal} {\bibinfo  {journal} {Science}\ }\textbf {\bibinfo
  {volume} {339}},\ \bibinfo {pages} {557} (\bibinfo {year} {2013})},\ \Eprint
  {http://arxiv.org/abs/arXiv:1312.2394v1} {arXiv:arXiv:1312.2394v1}
  \BibitemShut {NoStop}%
\bibitem [{\citenamefont {Wrachtrup}\ and\ \citenamefont
  {Jelezko}(2006)}]{Wrachtrup06rev}%
  \BibitemOpen
  \bibfield  {author} {\bibinfo {author} {\bibfnamefont {J.}~\bibnamefont
  {Wrachtrup}}\ and\ \bibinfo {author} {\bibfnamefont {F.}~\bibnamefont
  {Jelezko}},\ }\href@noop {} {\bibfield  {journal} {\bibinfo  {journal} {J.
  Phys.: Condens. Matter}\ }\textbf {\bibinfo {volume} {18}},\ \bibinfo {pages}
  {S807} (\bibinfo {year} {2006})}\BibitemShut {NoStop}%
\bibitem [{\citenamefont {Jelezko}\ \emph
  {et~al.}(2004{\natexlab{a}})\citenamefont {Jelezko}, \citenamefont {Gaebel},
  \citenamefont {Popa}, \citenamefont {Gruber},\ and\ \citenamefont
  {Wrachtrup}}]{Jelezko04a}%
  \BibitemOpen
  \bibfield  {author} {\bibinfo {author} {\bibfnamefont {F.}~\bibnamefont
  {Jelezko}}, \bibinfo {author} {\bibfnamefont {T.}~\bibnamefont {Gaebel}},
  \bibinfo {author} {\bibfnamefont {I.}~\bibnamefont {Popa}}, \bibinfo {author}
  {\bibfnamefont {A.}~\bibnamefont {Gruber}}, \ and\ \bibinfo {author}
  {\bibfnamefont {J.}~\bibnamefont {Wrachtrup}},\ }\href@noop {} {\bibfield
  {journal} {\bibinfo  {journal} {Phys. Rev. Lett.}\ }\textbf {\bibinfo
  {volume} {92}},\ \bibinfo {pages} {76401} (\bibinfo {year}
  {2004}{\natexlab{a}})}\BibitemShut {NoStop}%
\bibitem [{\citenamefont {Jelezko}\ \emph
  {et~al.}(2004{\natexlab{b}})\citenamefont {Jelezko}, \citenamefont {Gaebel},
  \citenamefont {Popa}, \citenamefont {Domhan}, \citenamefont {Gruber},\ and\
  \citenamefont {Wrachtrup}}]{Jelezko04b}%
  \BibitemOpen
  \bibfield  {author} {\bibinfo {author} {\bibfnamefont {F.}~\bibnamefont
  {Jelezko}}, \bibinfo {author} {\bibfnamefont {T.}~\bibnamefont {Gaebel}},
  \bibinfo {author} {\bibfnamefont {I.}~\bibnamefont {Popa}}, \bibinfo {author}
  {\bibfnamefont {M.}~\bibnamefont {Domhan}}, \bibinfo {author} {\bibfnamefont
  {A.}~\bibnamefont {Gruber}}, \ and\ \bibinfo {author} {\bibfnamefont
  {J.}~\bibnamefont {Wrachtrup}},\ }\href@noop {} {\bibfield  {journal}
  {\bibinfo  {journal} {Phys. Rev. Lett.}\ }\textbf {\bibinfo {volume} {93}},\
  \bibinfo {pages} {130501} (\bibinfo {year} {2004}{\natexlab{b}})}\BibitemShut
  {NoStop}%
\bibitem [{\citenamefont {Childress}\ \emph {et~al.}(2006)\citenamefont
  {Childress}, \citenamefont {Dutt}, \citenamefont {Taylor}, \citenamefont
  {Zibrov}, \citenamefont {Jelezko}, \citenamefont {Wrachtrup}, \citenamefont
  {Hemmer},\ and\ \citenamefont {Lukin}}]{Childress06b}%
  \BibitemOpen
  \bibfield  {author} {\bibinfo {author} {\bibfnamefont {L.}~\bibnamefont
  {Childress}}, \bibinfo {author} {\bibfnamefont {M.~V.~G.}\ \bibnamefont
  {Dutt}}, \bibinfo {author} {\bibfnamefont {J.~M.}\ \bibnamefont {Taylor}},
  \bibinfo {author} {\bibfnamefont {A.~S.}\ \bibnamefont {Zibrov}}, \bibinfo
  {author} {\bibfnamefont {F.}~\bibnamefont {Jelezko}}, \bibinfo {author}
  {\bibfnamefont {J.}~\bibnamefont {Wrachtrup}}, \bibinfo {author}
  {\bibfnamefont {P.~R.}\ \bibnamefont {Hemmer}}, \ and\ \bibinfo {author}
  {\bibfnamefont {M.~D.}\ \bibnamefont {Lukin}},\ }\href@noop {} {\bibfield
  {journal} {\bibinfo  {journal} {Science}\ }\textbf {\bibinfo {volume}
  {314}},\ \bibinfo {pages} {281} (\bibinfo {year} {2006})}\BibitemShut
  {NoStop}%
\bibitem [{\citenamefont {Balasubramanian}\ \emph {et~al.}(2009)\citenamefont
  {Balasubramanian}, \citenamefont {Neumann}, \citenamefont {Twitchen},
  \citenamefont {Markham}, \citenamefont {Kolesov}, \citenamefont {Mizuochi},
  \citenamefont {Isoya}, \citenamefont {Achard}, \citenamefont {Beck},
  \citenamefont {Tissler}, \citenamefont {Jacques}, \citenamefont {Hemmer},
  \citenamefont {Jelezko},\ and\ \citenamefont {Wrachtrup}}]{Balasub09}%
  \BibitemOpen
  \bibfield  {author} {\bibinfo {author} {\bibfnamefont {G.}~\bibnamefont
  {Balasubramanian}}, \bibinfo {author} {\bibfnamefont {P.}~\bibnamefont
  {Neumann}}, \bibinfo {author} {\bibfnamefont {D.}~\bibnamefont {Twitchen}},
  \bibinfo {author} {\bibfnamefont {M.}~\bibnamefont {Markham}}, \bibinfo
  {author} {\bibfnamefont {R.}~\bibnamefont {Kolesov}}, \bibinfo {author}
  {\bibfnamefont {N.}~\bibnamefont {Mizuochi}}, \bibinfo {author}
  {\bibfnamefont {J.}~\bibnamefont {Isoya}}, \bibinfo {author} {\bibfnamefont
  {J.}~\bibnamefont {Achard}}, \bibinfo {author} {\bibfnamefont
  {J.}~\bibnamefont {Beck}}, \bibinfo {author} {\bibfnamefont {J.}~\bibnamefont
  {Tissler}}, \bibinfo {author} {\bibfnamefont {V.}~\bibnamefont {Jacques}},
  \bibinfo {author} {\bibfnamefont {P.~R.}\ \bibnamefont {Hemmer}}, \bibinfo
  {author} {\bibfnamefont {F.}~\bibnamefont {Jelezko}}, \ and\ \bibinfo
  {author} {\bibfnamefont {J.}~\bibnamefont {Wrachtrup}},\ }\href@noop {}
  {\bibfield  {journal} {\bibinfo  {journal} {Nature Materials}\ }\textbf
  {\bibinfo {volume} {8}},\ \bibinfo {pages} {383} (\bibinfo {year}
  {2009})}\BibitemShut {NoStop}%
\bibitem [{\citenamefont {Doherty}\ \emph {et~al.}(2013)\citenamefont
  {Doherty}, \citenamefont {Manson}, \citenamefont {Delaney}, \citenamefont
  {Jelezko}, \citenamefont {Wrachtrup},\ and\ \citenamefont
  {Hollenberg}}]{Doherty2013}%
  \BibitemOpen
  \bibfield  {author} {\bibinfo {author} {\bibfnamefont {M.~W.}\ \bibnamefont
  {Doherty}}, \bibinfo {author} {\bibfnamefont {N.~B.}\ \bibnamefont {Manson}},
  \bibinfo {author} {\bibfnamefont {P.}~\bibnamefont {Delaney}}, \bibinfo
  {author} {\bibfnamefont {F.}~\bibnamefont {Jelezko}}, \bibinfo {author}
  {\bibfnamefont {J.}~\bibnamefont {Wrachtrup}}, \ and\ \bibinfo {author}
  {\bibfnamefont {L.~C.~L.}\ \bibnamefont {Hollenberg}},\ }\href {\doibase
  10.1016/j.physrep.2013.02.001} {\bibfield  {journal} {\bibinfo  {journal}
  {Physics Reports}\ }\textbf {\bibinfo {volume} {528}},\ \bibinfo {pages} {1}
  (\bibinfo {year} {2013})},\ \Eprint {http://arxiv.org/abs/1302.3288}
  {arXiv:1302.3288} \BibitemShut {NoStop}%
\bibitem [{\citenamefont {Shi}\ \emph {et~al.}(2018)\citenamefont {Shi},
  \citenamefont {Kong}, \citenamefont {Zhao}, \citenamefont {Zhang},
  \citenamefont {Chen}, \citenamefont {Chen}, \citenamefont {Zhang},
  \citenamefont {Wang}, \citenamefont {Ye}, \citenamefont {Wang}, \citenamefont
  {Qin}, \citenamefont {Rong}, \citenamefont {Su}, \citenamefont {Wang},
  \citenamefont {Qin},\ and\ \citenamefont {Du}}]{FShi:2018aa}%
  \BibitemOpen
  \bibfield  {author} {\bibinfo {author} {\bibfnamefont {F.}~\bibnamefont
  {Shi}}, \bibinfo {author} {\bibfnamefont {F.}~\bibnamefont {Kong}}, \bibinfo
  {author} {\bibfnamefont {P.}~\bibnamefont {Zhao}}, \bibinfo {author}
  {\bibfnamefont {X.}~\bibnamefont {Zhang}}, \bibinfo {author} {\bibfnamefont
  {M.}~\bibnamefont {Chen}}, \bibinfo {author} {\bibfnamefont {S.}~\bibnamefont
  {Chen}}, \bibinfo {author} {\bibfnamefont {Q.}~\bibnamefont {Zhang}},
  \bibinfo {author} {\bibfnamefont {M.}~\bibnamefont {Wang}}, \bibinfo {author}
  {\bibfnamefont {X.}~\bibnamefont {Ye}}, \bibinfo {author} {\bibfnamefont
  {Z.}~\bibnamefont {Wang}}, \bibinfo {author} {\bibfnamefont {Z.}~\bibnamefont
  {Qin}}, \bibinfo {author} {\bibfnamefont {X.}~\bibnamefont {Rong}}, \bibinfo
  {author} {\bibfnamefont {J.}~\bibnamefont {Su}}, \bibinfo {author}
  {\bibfnamefont {P.}~\bibnamefont {Wang}}, \bibinfo {author} {\bibfnamefont
  {P.~Z.}\ \bibnamefont {Qin}}, \ and\ \bibinfo {author} {\bibfnamefont
  {J.}~\bibnamefont {Du}},\ }\href@noop {} {\bibfield  {journal} {\bibinfo
  {journal} {Nat. Methods}\ }\textbf {\bibinfo {volume} {15}},\ \bibinfo
  {pages} {697} (\bibinfo {year} {2018})}\BibitemShut {NoStop}%
\bibitem [{\citenamefont {Fortman}\ and\ \citenamefont
  {Takahashi}(2019)}]{Fortman:2019aa}%
  \BibitemOpen
  \bibfield  {author} {\bibinfo {author} {\bibfnamefont {B.}~\bibnamefont
  {Fortman}}\ and\ \bibinfo {author} {\bibfnamefont {S.}~\bibnamefont
  {Takahashi}},\ }\href@noop {} {\bibfield  {journal} {\bibinfo  {journal} {J.
  Phys. Chem. A}\ }\textbf {\bibinfo {volume} {123}},\ \bibinfo {pages} {6350}
  (\bibinfo {year} {2019})}\BibitemShut {NoStop}%
\bibitem [{\citenamefont {Shi}\ \emph {et~al.}(2015)\citenamefont {Shi},
  \citenamefont {Zhang}, \citenamefont {Wang}, \citenamefont {Sun},
  \citenamefont {Wang}, \citenamefont {Rong}, \citenamefont {Chen},
  \citenamefont {Ju}, \citenamefont {Reinhard}, \citenamefont {Chen},
  \citenamefont {Wrachtrup}, \citenamefont {Wang},\ and\ \citenamefont
  {Du}}]{Shi:2015aa}%
  \BibitemOpen
  \bibfield  {author} {\bibinfo {author} {\bibfnamefont {F.}~\bibnamefont
  {Shi}}, \bibinfo {author} {\bibfnamefont {Q.}~\bibnamefont {Zhang}}, \bibinfo
  {author} {\bibfnamefont {P.}~\bibnamefont {Wang}}, \bibinfo {author}
  {\bibfnamefont {H.}~\bibnamefont {Sun}}, \bibinfo {author} {\bibfnamefont
  {J.}~\bibnamefont {Wang}}, \bibinfo {author} {\bibfnamefont {X.}~\bibnamefont
  {Rong}}, \bibinfo {author} {\bibfnamefont {M.}~\bibnamefont {Chen}}, \bibinfo
  {author} {\bibfnamefont {C.}~\bibnamefont {Ju}}, \bibinfo {author}
  {\bibfnamefont {F.}~\bibnamefont {Reinhard}}, \bibinfo {author}
  {\bibfnamefont {H.}~\bibnamefont {Chen}}, \bibinfo {author} {\bibfnamefont
  {J.}~\bibnamefont {Wrachtrup}}, \bibinfo {author} {\bibfnamefont
  {J.}~\bibnamefont {Wang}}, \ and\ \bibinfo {author} {\bibfnamefont
  {J.}~\bibnamefont {Du}},\ }\href {\doibase 10.1126/science.aaa2253}
  {\bibfield  {journal} {\bibinfo  {journal} {Science}\ }\textbf {\bibinfo
  {volume} {347}},\ \bibinfo {pages} {1135} (\bibinfo {year}
  {2015})}\BibitemShut {NoStop}%
\bibitem [{\citenamefont {Lovchinsky}\ \emph {et~al.}(2016)\citenamefont
  {Lovchinsky}, \citenamefont {Sushkov}, \citenamefont {Urbach}, \citenamefont
  {de~Leon}, \citenamefont {Choi}, \citenamefont {De~Greve}, \citenamefont
  {Evans}, \citenamefont {Gertner}, \citenamefont {Bersin}, \citenamefont
  {M{\"u}ller}, \citenamefont {McGuinness}, \citenamefont {Jelezko},
  \citenamefont {Walsworth}, \citenamefont {Park},\ and\ \citenamefont
  {Lukin}}]{Lovchinsky:2016bb}%
  \BibitemOpen
  \bibfield  {author} {\bibinfo {author} {\bibfnamefont {I.}~\bibnamefont
  {Lovchinsky}}, \bibinfo {author} {\bibfnamefont {A.~O.}\ \bibnamefont
  {Sushkov}}, \bibinfo {author} {\bibfnamefont {E.}~\bibnamefont {Urbach}},
  \bibinfo {author} {\bibfnamefont {N.~P.}\ \bibnamefont {de~Leon}}, \bibinfo
  {author} {\bibfnamefont {S.}~\bibnamefont {Choi}}, \bibinfo {author}
  {\bibfnamefont {K.}~\bibnamefont {De~Greve}}, \bibinfo {author}
  {\bibfnamefont {R.}~\bibnamefont {Evans}}, \bibinfo {author} {\bibfnamefont
  {R.}~\bibnamefont {Gertner}}, \bibinfo {author} {\bibfnamefont
  {E.}~\bibnamefont {Bersin}}, \bibinfo {author} {\bibfnamefont
  {C.}~\bibnamefont {M{\"u}ller}}, \bibinfo {author} {\bibfnamefont
  {L.}~\bibnamefont {McGuinness}}, \bibinfo {author} {\bibfnamefont
  {F.}~\bibnamefont {Jelezko}}, \bibinfo {author} {\bibfnamefont {R.~L.}\
  \bibnamefont {Walsworth}}, \bibinfo {author} {\bibfnamefont {H.}~\bibnamefont
  {Park}}, \ and\ \bibinfo {author} {\bibfnamefont {M.~D.}\ \bibnamefont
  {Lukin}},\ }\href {\doibase 10.1126/science.aad8022} {\bibfield  {journal}
  {\bibinfo  {journal} {Science}\ }\textbf {\bibinfo {volume} {351}},\ \bibinfo
  {pages} {836} (\bibinfo {year} {2016})}\BibitemShut {NoStop}%
\bibitem [{\citenamefont {Akiel}\ \emph {et~al.}(2016)\citenamefont {Akiel},
  \citenamefont {Zhang}, \citenamefont {Abeywardana}, \citenamefont {Stepanov},
  \citenamefont {Qin},\ and\ \citenamefont {Takahashi}}]{Akiel:2016aa}%
  \BibitemOpen
  \bibfield  {author} {\bibinfo {author} {\bibfnamefont {R.~D.}\ \bibnamefont
  {Akiel}}, \bibinfo {author} {\bibfnamefont {X.}~\bibnamefont {Zhang}},
  \bibinfo {author} {\bibfnamefont {C.}~\bibnamefont {Abeywardana}}, \bibinfo
  {author} {\bibfnamefont {V.}~\bibnamefont {Stepanov}}, \bibinfo {author}
  {\bibfnamefont {P.~Z.}\ \bibnamefont {Qin}}, \ and\ \bibinfo {author}
  {\bibfnamefont {S.}~\bibnamefont {Takahashi}},\ }\bibfield  {booktitle}
  {\emph {\bibinfo {booktitle} {The Journal of Physical Chemistry B}},\ }\href
  {\doibase 10.1021/acs.jpcb.6b00790} {\bibfield  {journal} {\bibinfo
  {journal} {The Journal of Physical Chemistry B}\ }\textbf {\bibinfo {volume}
  {120}},\ \bibinfo {pages} {4003} (\bibinfo {year} {2016})}\BibitemShut
  {NoStop}%
\bibitem [{\citenamefont {Akiel}\ \emph {et~al.}(2017)\citenamefont {Akiel},
  \citenamefont {Stepanov},\ and\ \citenamefont {Takahashi}}]{Akiel:2017aa}%
  \BibitemOpen
  \bibfield  {author} {\bibinfo {author} {\bibfnamefont {R.~D.}\ \bibnamefont
  {Akiel}}, \bibinfo {author} {\bibfnamefont {V.}~\bibnamefont {Stepanov}}, \
  and\ \bibinfo {author} {\bibfnamefont {S.}~\bibnamefont {Takahashi}},\ }\href
  {\doibase 10.1007/s12013-016-0739-4} {\bibfield  {journal} {\bibinfo
  {journal} {Cell Biochemistry and Biophysics}\ }\textbf {\bibinfo {volume}
  {75}},\ \bibinfo {pages} {151} (\bibinfo {year} {2017})}\BibitemShut
  {NoStop}%
\bibitem [{\citenamefont {Cunningham}\ \emph {et~al.}(2015)\citenamefont
  {Cunningham}, \citenamefont {Putterman}, \citenamefont {Desai}, \citenamefont
  {Horne},\ and\ \citenamefont {Saxena}}]{Cunningham:2015aa}%
  \BibitemOpen
  \bibfield  {author} {\bibinfo {author} {\bibfnamefont {T.~F.}\ \bibnamefont
  {Cunningham}}, \bibinfo {author} {\bibfnamefont {M.~R.}\ \bibnamefont
  {Putterman}}, \bibinfo {author} {\bibfnamefont {A.}~\bibnamefont {Desai}},
  \bibinfo {author} {\bibfnamefont {W.~S.}\ \bibnamefont {Horne}}, \ and\
  \bibinfo {author} {\bibfnamefont {S.}~\bibnamefont {Saxena}},\ }\href
  {\doibase 10.1002/anie.201501968} {\bibfield  {journal} {\bibinfo  {journal}
  {Angewandte Chemie (International ed. in English)}\ }\textbf {\bibinfo
  {volume} {54}},\ \bibinfo {pages} {6330} (\bibinfo {year}
  {2015})}\BibitemShut {NoStop}%
\bibitem [{\citenamefont {Sameach}\ \emph {et~al.}(2019)\citenamefont
  {Sameach}, \citenamefont {Ghosh}, \citenamefont {Gevorkyan-Airapetov},
  \citenamefont {Saxena},\ and\ \citenamefont {Ruthstein}}]{Sameach:2019aa}%
  \BibitemOpen
  \bibfield  {author} {\bibinfo {author} {\bibfnamefont {H.}~\bibnamefont
  {Sameach}}, \bibinfo {author} {\bibfnamefont {S.}~\bibnamefont {Ghosh}},
  \bibinfo {author} {\bibfnamefont {L.}~\bibnamefont {Gevorkyan-Airapetov}},
  \bibinfo {author} {\bibfnamefont {S.}~\bibnamefont {Saxena}}, \ and\ \bibinfo
  {author} {\bibfnamefont {S.}~\bibnamefont {Ruthstein}},\ }\href@noop {}
  {\bibfield  {journal} {\bibinfo  {journal} {Agnew. Chemie Int. Ed.}\ }\textbf
  {\bibinfo {volume} {58}},\ \bibinfo {pages} {3053} (\bibinfo {year}
  {2019})}\BibitemShut {NoStop}%
\bibitem [{\citenamefont {Lawless}\ \emph {et~al.}(2018)\citenamefont
  {Lawless}, \citenamefont {Pettersson}, \citenamefont {Rule}, \citenamefont
  {Lanni},\ and\ \citenamefont {Saxena}}]{Lawless:2018aa}%
  \BibitemOpen
  \bibfield  {author} {\bibinfo {author} {\bibfnamefont {M.~J.}\ \bibnamefont
  {Lawless}}, \bibinfo {author} {\bibfnamefont {J.~R.}\ \bibnamefont
  {Pettersson}}, \bibinfo {author} {\bibfnamefont {G.~S.}\ \bibnamefont
  {Rule}}, \bibinfo {author} {\bibfnamefont {F.}~\bibnamefont {Lanni}}, \ and\
  \bibinfo {author} {\bibfnamefont {S.}~\bibnamefont {Saxena}},\ }\href@noop {}
  {\bibfield  {journal} {\bibinfo  {journal} {Biophys. J.}\ }\textbf {\bibinfo
  {volume} {114}},\ \bibinfo {pages} {592} (\bibinfo {year}
  {2018})}\BibitemShut {NoStop}%
\bibitem [{\citenamefont {Singewald}\ \emph {et~al.}(2020)\citenamefont
  {Singewald}, \citenamefont {Bogetti}, \citenamefont {Sinha}, \citenamefont
  {Rule},\ and\ \citenamefont {Saxena}}]{Singewald:2020aa}%
  \BibitemOpen
  \bibfield  {author} {\bibinfo {author} {\bibfnamefont {K.}~\bibnamefont
  {Singewald}}, \bibinfo {author} {\bibfnamefont {X.}~\bibnamefont {Bogetti}},
  \bibinfo {author} {\bibfnamefont {K.}~\bibnamefont {Sinha}}, \bibinfo
  {author} {\bibfnamefont {G.~S.}\ \bibnamefont {Rule}}, \ and\ \bibinfo
  {author} {\bibfnamefont {S.}~\bibnamefont {Saxena}},\ }\href {\doibase
  https://doi.org/10.1002/anie.202009982} {\bibfield  {journal} {\bibinfo
  {journal} {Angewandte Chemie International Edition}\ }\textbf {\bibinfo
  {volume} {59}},\ \bibinfo {pages} {23040} (\bibinfo {year}
  {2020})}\BibitemShut {NoStop}%
\bibitem [{\citenamefont {Ghosh}\ \emph {et~al.}(2018)\citenamefont {Ghosh},
  \citenamefont {Lawless}, \citenamefont {Rule},\ and\ \citenamefont
  {Saxena}}]{Ghosh:2018aa}%
  \BibitemOpen
  \bibfield  {author} {\bibinfo {author} {\bibfnamefont {S.}~\bibnamefont
  {Ghosh}}, \bibinfo {author} {\bibfnamefont {M.~J.}\ \bibnamefont {Lawless}},
  \bibinfo {author} {\bibfnamefont {G.~S.}\ \bibnamefont {Rule}}, \ and\
  \bibinfo {author} {\bibfnamefont {S.}~\bibnamefont {Saxena}},\ }\href@noop {}
  {\bibfield  {journal} {\bibinfo  {journal} {J. Magn. Res.}\ }\textbf
  {\bibinfo {volume} {286}},\ \bibinfo {pages} {163} (\bibinfo {year}
  {2018})}\BibitemShut {NoStop}%
\bibitem [{\citenamefont {{Gamble Jarvi}}\ \emph {et~al.}(2018)\citenamefont
  {{Gamble Jarvi}}, \citenamefont {Ranguelova}, \citenamefont {Ghosh},
  \citenamefont {Weber},\ and\ \citenamefont {Saxena}}]{Jarvi:2018aa}%
  \BibitemOpen
  \bibfield  {author} {\bibinfo {author} {\bibfnamefont {A.}~\bibnamefont
  {{Gamble Jarvi}}}, \bibinfo {author} {\bibfnamefont {K.}~\bibnamefont
  {Ranguelova}}, \bibinfo {author} {\bibfnamefont {S.}~\bibnamefont {Ghosh}},
  \bibinfo {author} {\bibfnamefont {R.~T.}\ \bibnamefont {Weber}}, \ and\
  \bibinfo {author} {\bibfnamefont {S.}~\bibnamefont {Saxena}},\ }\href@noop {}
  {\bibfield  {journal} {\bibinfo  {journal} {J. Phys. Chem. B}\ }\textbf
  {\bibinfo {volume} {122}},\ \bibinfo {pages} {10669} (\bibinfo {year}
  {2018})}\BibitemShut {NoStop}%
\bibitem [{\citenamefont {Bogetti}\ \emph {et~al.}(2020)\citenamefont
  {Bogetti}, \citenamefont {Ghosh}, \citenamefont {{Gamble Jarvi}},
  \citenamefont {Wang},\ and\ \citenamefont {Saxena}}]{Bogetti:2020aa}%
  \BibitemOpen
  \bibfield  {author} {\bibinfo {author} {\bibfnamefont {X.}~\bibnamefont
  {Bogetti}}, \bibinfo {author} {\bibfnamefont {S.}~\bibnamefont {Ghosh}},
  \bibinfo {author} {\bibfnamefont {A.}~\bibnamefont {{Gamble Jarvi}}},
  \bibinfo {author} {\bibfnamefont {J.}~\bibnamefont {Wang}}, \ and\ \bibinfo
  {author} {\bibfnamefont {S.}~\bibnamefont {Saxena}},\ }\href@noop {}
  {\bibfield  {journal} {\bibinfo  {journal} {J. Phys. Chem. B}\ }\textbf
  {\bibinfo {volume} {124}},\ \bibinfo {pages} {2788} (\bibinfo {year}
  {2020})}\BibitemShut {NoStop}%
\bibitem [{\citenamefont {Grotz}\ \emph {et~al.}(2011)\citenamefont {Grotz},
  \citenamefont {Beck}, \citenamefont {Neumann}, \citenamefont {Naydenov},
  \citenamefont {Reuter}, \citenamefont {Reinhard}, \citenamefont {Jelezko},
  \citenamefont {Wrachtrup}, \citenamefont {Schweinfurth}, \citenamefont
  {Sarkar},\ and\ \citenamefont {Hemmer}}]{Grotz2011}%
  \BibitemOpen
  \bibfield  {author} {\bibinfo {author} {\bibfnamefont {B.}~\bibnamefont
  {Grotz}}, \bibinfo {author} {\bibfnamefont {J.}~\bibnamefont {Beck}},
  \bibinfo {author} {\bibfnamefont {P.}~\bibnamefont {Neumann}}, \bibinfo
  {author} {\bibfnamefont {B.}~\bibnamefont {Naydenov}}, \bibinfo {author}
  {\bibfnamefont {R.}~\bibnamefont {Reuter}}, \bibinfo {author} {\bibfnamefont
  {F.}~\bibnamefont {Reinhard}}, \bibinfo {author} {\bibfnamefont
  {F.}~\bibnamefont {Jelezko}}, \bibinfo {author} {\bibfnamefont
  {J.}~\bibnamefont {Wrachtrup}}, \bibinfo {author} {\bibfnamefont
  {D.}~\bibnamefont {Schweinfurth}}, \bibinfo {author} {\bibfnamefont
  {B.}~\bibnamefont {Sarkar}}, \ and\ \bibinfo {author} {\bibfnamefont
  {P.}~\bibnamefont {Hemmer}},\ }\href {\doibase 10.1088/1367-2630/13/5/055004}
  {\bibfield  {journal} {\bibinfo  {journal} {New Journal of Physics}\ }\textbf
  {\bibinfo {volume} {13}} (\bibinfo {year} {2011}),\
  10.1088/1367-2630/13/5/055004}\BibitemShut {NoStop}%
\bibitem [{\citenamefont {Mamin}\ \emph {et~al.}(2012)\citenamefont {Mamin},
  \citenamefont {Sherwood},\ and\ \citenamefont {Rugar}}]{Mamin2012:bb}%
  \BibitemOpen
  \bibfield  {author} {\bibinfo {author} {\bibfnamefont {H.~J.}\ \bibnamefont
  {Mamin}}, \bibinfo {author} {\bibfnamefont {M.~H.}\ \bibnamefont {Sherwood}},
  \ and\ \bibinfo {author} {\bibfnamefont {D.}~\bibnamefont {Rugar}},\ }\href
  {\doibase 10.1103/PhysRevB.86.195422} {\bibfield  {journal} {\bibinfo
  {journal} {Phys. Rev. B}\ }\textbf {\bibinfo {volume} {86}},\ \bibinfo
  {pages} {195422} (\bibinfo {year} {2012})}\BibitemShut {NoStop}%
\bibitem [{\citenamefont {Sushkov}\ \emph {et~al.}(2014)\citenamefont
  {Sushkov}, \citenamefont {Lovchinsky}, \citenamefont {Chisholm},
  \citenamefont {Walsworth}, \citenamefont {Park},\ and\ \citenamefont
  {Lukin}}]{Sushkov2014a}%
  \BibitemOpen
  \bibfield  {author} {\bibinfo {author} {\bibfnamefont {A.~O.}\ \bibnamefont
  {Sushkov}}, \bibinfo {author} {\bibfnamefont {I.}~\bibnamefont {Lovchinsky}},
  \bibinfo {author} {\bibfnamefont {N.}~\bibnamefont {Chisholm}}, \bibinfo
  {author} {\bibfnamefont {R.~L.}\ \bibnamefont {Walsworth}}, \bibinfo {author}
  {\bibfnamefont {H.}~\bibnamefont {Park}}, \ and\ \bibinfo {author}
  {\bibfnamefont {M.~D.}\ \bibnamefont {Lukin}},\ }\href {\doibase
  10.1103/PhysRevLett.113.197601} {\bibfield  {journal} {\bibinfo  {journal}
  {Physical Review Letters}\ }\textbf {\bibinfo {volume} {113}},\ \bibinfo
  {pages} {1} (\bibinfo {year} {2014})},\ \Eprint
  {http://arxiv.org/abs/1410.3750} {arXiv:1410.3750} \BibitemShut {NoStop}%
\bibitem [{\citenamefont {Grinolds}\ \emph {et~al.}(2014)\citenamefont
  {Grinolds}, \citenamefont {Warner}, \citenamefont {{De Greve}}, \citenamefont
  {Dovzhenko}, \citenamefont {Thiel}, \citenamefont {Walsworth}, \citenamefont
  {Hong}, \citenamefont {Maletinsky},\ and\ \citenamefont
  {Yacoby}}]{Grinolds2014}%
  \BibitemOpen
  \bibfield  {author} {\bibinfo {author} {\bibfnamefont {M.~S.}\ \bibnamefont
  {Grinolds}}, \bibinfo {author} {\bibfnamefont {M.}~\bibnamefont {Warner}},
  \bibinfo {author} {\bibfnamefont {K.}~\bibnamefont {{De Greve}}}, \bibinfo
  {author} {\bibfnamefont {Y.}~\bibnamefont {Dovzhenko}}, \bibinfo {author}
  {\bibfnamefont {L.}~\bibnamefont {Thiel}}, \bibinfo {author} {\bibfnamefont
  {R.~L.}\ \bibnamefont {Walsworth}}, \bibinfo {author} {\bibfnamefont
  {S.}~\bibnamefont {Hong}}, \bibinfo {author} {\bibfnamefont {P.}~\bibnamefont
  {Maletinsky}}, \ and\ \bibinfo {author} {\bibfnamefont {a.}~\bibnamefont
  {Yacoby}},\ }\href {\doibase 10.1038/nnano.2014.30} {\bibfield  {journal}
  {\bibinfo  {journal} {Nature nanotechnology}\ }\textbf {\bibinfo {volume}
  {9}},\ \bibinfo {pages} {279} (\bibinfo {year} {2014})},\ \Eprint
  {http://arxiv.org/abs/1401.2674} {arXiv:1401.2674} \BibitemShut {NoStop}%
\bibitem [{\citenamefont {Zhang}(2018)}]{KaiThesis2018}%
  \BibitemOpen
  \bibfield  {author} {\bibinfo {author} {\bibfnamefont {K.}~\bibnamefont
  {Zhang}},\ }\emph {\bibinfo {title} {EXPLORING NANOSCALE SPIN PHYSICS USING
  SINGLE SPINS IN DIAMOND}},\ \href {http://d-scholarship.pitt.edu/35020/}
  {Ph.D. thesis},\ \bibinfo  {school} {University of Pittsburgh} (\bibinfo
  {year} {2018})\BibitemShut {NoStop}%
\bibitem [{\citenamefont {Abeywardana}\ \emph {et~al.}(2016)\citenamefont
  {Abeywardana}, \citenamefont {Stepanov}, \citenamefont {Cho},\ and\
  \citenamefont {Takahashi}}]{Abeywardana:2016aa}%
  \BibitemOpen
  \bibfield  {author} {\bibinfo {author} {\bibfnamefont {C.}~\bibnamefont
  {Abeywardana}}, \bibinfo {author} {\bibfnamefont {V.}~\bibnamefont
  {Stepanov}}, \bibinfo {author} {\bibfnamefont {F.~H.}\ \bibnamefont {Cho}}, \
  and\ \bibinfo {author} {\bibfnamefont {S.}~\bibnamefont {Takahashi}},\ }\href
  {\doibase 10.1063/1.4963717} {\bibfield  {journal} {\bibinfo  {journal}
  {Journal of Applied Physics}\ }\textbf {\bibinfo {volume} {120}},\ \bibinfo
  {pages} {123907} (\bibinfo {year} {2016})},\ \Eprint
  {http://arxiv.org/abs/https://doi.org/10.1063/1.4963717}
  {https://doi.org/10.1063/1.4963717} \BibitemShut {NoStop}%
\bibitem [{\citenamefont {Stepanov}\ and\ \citenamefont
  {Takahashi}(2016)}]{Stepanov:2016aa}%
  \BibitemOpen
  \bibfield  {author} {\bibinfo {author} {\bibfnamefont {V.}~\bibnamefont
  {Stepanov}}\ and\ \bibinfo {author} {\bibfnamefont {S.}~\bibnamefont
  {Takahashi}},\ }\href {\doibase 10.1103/PhysRevB.94.024421} {\bibfield
  {journal} {\bibinfo  {journal} {Physical Review B}\ }\textbf {\bibinfo
  {volume} {94}},\ \bibinfo {pages} {1} (\bibinfo {year} {2016})},\ \bibinfo
  {note} {\_eprint: 1603.07404v1}\BibitemShut {NoStop}%
\bibitem [{\citenamefont {Yang}\ \emph {et~al.}(2015)\citenamefont {Yang},
  \citenamefont {Ji}, \citenamefont {Cunningham},\ and\ \citenamefont
  {Saxena}}]{Yang2015:bb}%
  \BibitemOpen
  \bibfield  {author} {\bibinfo {author} {\bibfnamefont {Z.}~\bibnamefont
  {Yang}}, \bibinfo {author} {\bibfnamefont {M.}~\bibnamefont {Ji}}, \bibinfo
  {author} {\bibfnamefont {T.~F.}\ \bibnamefont {Cunningham}}, \ and\ \bibinfo
  {author} {\bibfnamefont {S.}~\bibnamefont {Saxena}},\ }in\ \href {\doibase
  https://doi.org/10.1016/bs.mie.2015.05.026} {\emph {\bibinfo {booktitle}
  {Electron Paramagnetic Resonance Investigations of Biological Systems by
  Using Spin Labels, Spin Probes, and Intrinsic Metal Ions, Part A}}},\
  \bibinfo {series} {Methods in Enzymology}, Vol.\ \bibinfo {volume} {563},\
  \bibinfo {editor} {edited by\ \bibinfo {editor} {\bibfnamefont {P.~Z.}\
  \bibnamefont {Qin}}\ and\ \bibinfo {editor} {\bibfnamefont {K.}~\bibnamefont
  {Warncke}}}\ (\bibinfo  {publisher} {Academic Press},\ \bibinfo {year}
  {2015})\ pp.\ \bibinfo {pages} {459 -- 481}\BibitemShut {NoStop}%
\bibitem [{\citenamefont {Ji}\ \emph {et~al.}(2014)\citenamefont {Ji},
  \citenamefont {Ruthstein},\ and\ \citenamefont {Saxena}}]{Ji:2014aa}%
  \BibitemOpen
  \bibfield  {author} {\bibinfo {author} {\bibfnamefont {M.}~\bibnamefont
  {Ji}}, \bibinfo {author} {\bibfnamefont {S.}~\bibnamefont {Ruthstein}}, \
  and\ \bibinfo {author} {\bibfnamefont {S.}~\bibnamefont {Saxena}},\
  }\bibfield  {booktitle} {\emph {\bibinfo {booktitle} {Accounts of Chemical
  Research}},\ }\href {\doibase 10.1021/ar400245z} {\bibfield  {journal}
  {\bibinfo  {journal} {Accounts of Chemical Research}\ }\textbf {\bibinfo
  {volume} {47}},\ \bibinfo {pages} {688} (\bibinfo {year} {2014})}\BibitemShut
  {NoStop}%
\bibitem [{\citenamefont {Santana}\ \emph {et~al.}(2005)\citenamefont
  {Santana}, \citenamefont {Cunha}, \citenamefont {Carvalho}, \citenamefont
  {Vencato},\ and\ \citenamefont {Calvo}}]{santana_single_2005}%
  \BibitemOpen
  \bibfield  {author} {\bibinfo {author} {\bibfnamefont {R.~C.}\ \bibnamefont
  {Santana}}, \bibinfo {author} {\bibfnamefont {R.~O.}\ \bibnamefont {Cunha}},
  \bibinfo {author} {\bibfnamefont {J.~F.}\ \bibnamefont {Carvalho}}, \bibinfo
  {author} {\bibfnamefont {I.}~\bibnamefont {Vencato}}, \ and\ \bibinfo
  {author} {\bibfnamefont {R.}~\bibnamefont {Calvo}},\ }\href {\doibase
  10.1016/j.jinorgbio.2004.10.014} {\bibfield  {journal} {\bibinfo  {journal}
  {Journal of Inorganic Biochemistry}\ }\textbf {\bibinfo {volume} {99}},\
  \bibinfo {pages} {415} (\bibinfo {year} {2005})}\BibitemShut {NoStop}%
\bibitem [{\citenamefont {Neuman}\ \emph {et~al.}(2010)\citenamefont {Neuman},
  \citenamefont {Perec}, \citenamefont {Gonzalez}, \citenamefont {Passeggi},
  \citenamefont {Rizzi},\ and\ \citenamefont {Brondino}}]{neuman_single_2010}%
  \BibitemOpen
  \bibfield  {author} {\bibinfo {author} {\bibfnamefont {N.~I.}\ \bibnamefont
  {Neuman}}, \bibinfo {author} {\bibfnamefont {M.}~\bibnamefont {Perec}},
  \bibinfo {author} {\bibfnamefont {P.~J.}\ \bibnamefont {Gonzalez}}, \bibinfo
  {author} {\bibfnamefont {M.~C.~G.}\ \bibnamefont {Passeggi}}, \bibinfo
  {author} {\bibfnamefont {A.~C.}\ \bibnamefont {Rizzi}}, \ and\ \bibinfo
  {author} {\bibfnamefont {C.~D.}\ \bibnamefont {Brondino}},\ }\href {\doibase
  10.1021/jp108736p} {\bibfield  {journal} {\bibinfo  {journal} {The Journal of
  Physical Chemistry A}\ }\textbf {\bibinfo {volume} {114}},\ \bibinfo {pages}
  {13069} (\bibinfo {year} {2010})},\ \bibinfo {note} {publisher: American
  Chemical Society}\BibitemShut {NoStop}%
\bibitem [{\citenamefont {Neuman}\ \emph {et~al.}(2012)\citenamefont {Neuman},
  \citenamefont {Franco}, \citenamefont {Ferroni}, \citenamefont {Baggio},
  \citenamefont {Passeggi}, \citenamefont {Rizzi},\ and\ \citenamefont
  {Brondino}}]{neuman_single_2012}%
  \BibitemOpen
  \bibfield  {author} {\bibinfo {author} {\bibfnamefont {N.~I.}\ \bibnamefont
  {Neuman}}, \bibinfo {author} {\bibfnamefont {V.~G.}\ \bibnamefont {Franco}},
  \bibinfo {author} {\bibfnamefont {F.~M.}\ \bibnamefont {Ferroni}}, \bibinfo
  {author} {\bibfnamefont {R.}~\bibnamefont {Baggio}}, \bibinfo {author}
  {\bibfnamefont {M.~C.~G.}\ \bibnamefont {Passeggi}}, \bibinfo {author}
  {\bibfnamefont {A.~C.}\ \bibnamefont {Rizzi}}, \ and\ \bibinfo {author}
  {\bibfnamefont {C.~D.}\ \bibnamefont {Brondino}},\ }\href {\doibase
  10.1021/jp308745e} {\bibfield  {journal} {\bibinfo  {journal} {J. Phys. Chem.
  A}\ }\textbf {\bibinfo {volume} {116}},\ \bibinfo {pages} {12314} (\bibinfo
  {year} {2012})},\ \bibinfo {note} {archive Location: world Publisher:
  American Chemical Society}\BibitemShut {NoStop}%
\bibitem [{\citenamefont {Stoll}\ and\ \citenamefont
  {Schweiger}(2006)}]{Stoll:2006aa}%
  \BibitemOpen
  \bibfield  {author} {\bibinfo {author} {\bibfnamefont {S.}~\bibnamefont
  {Stoll}}\ and\ \bibinfo {author} {\bibfnamefont {A.}~\bibnamefont
  {Schweiger}},\ }\href {\doibase 10.1016/j.jmr.2005.08.013} {\bibfield
  {journal} {\bibinfo  {journal} {J Magn Reson}\ }\textbf {\bibinfo {volume}
  {178}},\ \bibinfo {pages} {42} (\bibinfo {year} {2006})}\BibitemShut
  {NoStop}%
\bibitem [{\citenamefont {Kolkowitz}\ \emph {et~al.}(2012)\citenamefont
  {Kolkowitz}, \citenamefont {Unterreithmeier}, \citenamefont {Bennett},\ and\
  \citenamefont {Lukin}}]{Kolkowitz2012distantnucspin}%
  \BibitemOpen
  \bibfield  {author} {\bibinfo {author} {\bibfnamefont {S.}~\bibnamefont
  {Kolkowitz}}, \bibinfo {author} {\bibfnamefont {Q.~P.}\ \bibnamefont
  {Unterreithmeier}}, \bibinfo {author} {\bibfnamefont {S.~D.}\ \bibnamefont
  {Bennett}}, \ and\ \bibinfo {author} {\bibfnamefont {M.~D.}\ \bibnamefont
  {Lukin}},\ }\href {\doibase 10.1103/PhysRevLett.109.137601} {\bibfield
  {journal} {\bibinfo  {journal} {Phys. Rev. Lett.}\ }\textbf {\bibinfo
  {volume} {109}},\ \bibinfo {pages} {137601} (\bibinfo {year}
  {2012})}\BibitemShut {NoStop}%
\end{thebibliography}
%

\newpage

\newpage

\newpage

\newpage

\newpage

\newpage

\newpage

\newpage

\newpage

\end{document}